\documentclass[useAMS,usenatbib,a4paper,fleqn]{mn2e}

\usepackage{amsfonts,amssymb,amsmath}
\usepackage{aas_macros}
\usepackage{times,varioref,multirow,textcomp,comment}
\usepackage[usenames,dvipsnames,svgnames,hyperref]{xcolor}
\usepackage[pdftex]{graphicx}
\usepackage{comment}
\usepackage[
    pdftex,% get better results and can break links but can't use psfrag
    a4paper=false,
    plainpages=true,
    pdfpagelabels,
    breaklinks=true,
    bookmarks=true,
    bookmarksopen=false,
    bookmarksopenlevel=2
    bookmarksnumbered=true,
    bookmarkstype=toc,
    colorlinks=true,
    citecolor=RoyalBlue,
    linkcolor=ForestGreen,
    menucolor=Teal,
    urlcolor=DarkOrange,
]{hyperref}

\providecommand{\adsurl}[1]{\href{#1}{ADS}}

\hypersetup{
  pdfauthor = {Pascal Elahi, Hareth Mahdi, Chris Power, Geraint Lewis},
  pdfkeywords = {Lensing, Warm Dark Matter},
  pdftitle = {}
}

\def \Msun{\ {\rm M_\odot}}
\def \Msunh{\ h^{-1}{\rm M_\odot}}

\def \Mpch{\ h^{-1}{\rm Mpc}}

\def \LCDM{$\Lambda$CDM}
\def \rvmax{R_{V_{\rm max}}}
\def \vmax{V_{\rm max}}
\def \rnfw{r_{\rm NFW}}

\newcommand{\Eqref}[1]{Eq.~(\ref{#1})}
\newcommand{\Figref}[1]{Fig.~\ref{#1}}
\newcommand{\Secref}[1]{\S\ref{#1}}

\defcitealias{nfw}{NFW}
\def \nfw{NFW}
\defcitealias{elahi2011}{{\sc stf}}
\def \stf{{\sc stf}}
\def \velociraptor{{\sc veloci}raptor}
\defcitealias{ahf}{{\sc ahf}}
\def \ahf{{\sc ahf}}
\def \Gadget2{{\sc gadget-2}}

\begin{document}
\title[WDM Mass Accretion \& Lensing]{Warm Dark Haloes Accretion Histories and their Gravitational Signatures}
\author[P.J.~Elahi, {\it et al}.]{
Pascal J. Elahi\thanks{E-mail: pelahi@physics.usyd.edu.au},$^{1}$
Hareth S. Mahdi,$^{1,2}$
Chris Power$^{3}$,
Geraint F. Lewis,$^{1}$
\\
%\parbox{\textwidth}{
$^{1}$Sydney Institute for Astronomy, A28, School of Physics, The University of Sydney, NSW 2006, Australia\\
$^{2}$Department of Astronomy, University of Baghdad, Jadiryah, Baghdad 10071, Iraq \\
$^{3}$International Centre for Radio Astronomy Research, University of Western Australia, 35 Stirling Highway, Crawley, WA 6009, Australia\\
%}
}
\maketitle

\pdfbookmark[1]{Abstract}{sec:abstract}
\begin{abstract}
We study clusters in Warm Dark Matter (WDM) models of a thermally produced dark matter particle $0.5$~keV in mass. We show that, despite clusters in WDM cosmologies having similar density profiles as their Cold Dark Matter (CDM) counterparts, the internal properties, such as the amount of substructure, shows marked differences. This result is surprising as clusters are at mass scales that are {\em a thousand times greater} than that at which structure formation is suppressed. WDM clusters gain significantly more mass via smooth accretion and contain fewer substructures than their CDM brethren. The higher smooth mass accretion results in subhaloes which are physically more extended and less dense. These fine-scale differences can be probed by strong gravitational lensing. We find, unexpectedly, that WDM clusters have {\em higher} lensing efficiencies than those in CDM cosmologies, contrary to the naive expectation that WDM clusters should be less efficient due to the fewer substructures they contain. Despite being less dense, the larger WDM subhaloes are more likely to have larger lensing cross-sections than CDM ones. Additionally, WDM subhaloes typically reside at larger distances, which radially stretches the critical lines associated with strong gravitational lensing, resulting in excess in the number of clusters with large radial cross-sections at the $\sim2\sigma$ level. Though lensing profile for an individual cluster vary significantly with the line-of-sight, the radial arc distribution based on a sample of $\gtrsim100$ clusters may prove to be the crucial test for the presence of WDM.
\end{abstract}
\maketitle

%---------------------------------
\section{Introduction}\label{sec:intro}
%this intro and the references are based in part from lovell 2013 wdm 
The nature of Dark Matter (DM) remains one of the central mysteries in cosmology. Various lines of evidence, such as cosmic microwave background, indicate that this matter is composed of nonbaryonic elementary particles \cite[e.g.][]{wmap9}, though exactly what type of DM is not yet known. Much of the focus has been on so-called Cold Dark Matter (CDM) \cite[see][for a review]{frenkwhite2012}, for which there are well-motivated candidates from particle physics, {\it e.g.} the lightest supersymmetric particle or neutralino \cite[][]{ellis1984}, or the axion \cite[][]{preskill1983} \cite[for a summary of several candidates, see for instance][]{bertone2005,petraki2013}. The key feature of all of these candidates is that the particles have {\em negligible} thermal velocities during the era of structure formation. 

\par
Though it often claimed that cosmological data favours cold dark matter, it would be more accurate to say that it rules out hot dark matter, where the particle, such as the ordinary neutrino, became non-relativistic around the time of recombination \cite[e.g.][]{fof}. Tensions between observations and predictions from CDM models do exist on galactic scales and has renewed interest in other types of dark matter, such as {\em Warm} Dark Matter (WDM) \cite[e.g.][]{schneider2012,lovell2012,libeskind2013}. WDM models have DM particles with appreciable but non-relativistic thermal velocities at early times. The best-known example is a sterile neutrino, the so-called neutrino minimal standard model, which could explain observed neutrino oscillation rates and baryogenesis \cite[$\nu$MS; e.g.][]{asaka2006}.

\par
Due to the thermal velocities, particles free stream out of scale-small perturbations, giving rise to a cutoff in the linear matter power spectrum and an associated suppression of structure formation at and below these scales. These models have the benefit of reproducing the observed large-scale matter distribution while possibly resolving tensions that exist on smaller scales in CDM models. One of the most well known problem with CDM models is the so-called ``missing satellite problem'': CDM models predict many more satellite galaxies than observed around galaxies such as our own \cite[e.g.][]{klypin1999,moore1999}. The excess number of dark matter subhaloes that should host satellite galaxies can simply mean that these substructures did not confine gas and are therefore completely dark \cite[e.g.][]{bullock2000,benson2002,nickerson2011,nickerson2012}. However, numerical simulations show that CDM models invariably predict several satellites that are “too big” to be masked by galaxy formation processes, at odds with observations \cite[e.g.][]{boylankolchin2011,boylankolchin2012}. \cite{lovell2012} showed that the resonantly produced sterile neutrino DM models, compatible with the Lyman-$\alpha$ bounds \cite[][]{boyarsky2009a,boyarsky2009b}, decrease the number of substructures residing in a Galaxy-size halo. Certain WDM particle candidates have masses that result in the suppression of growth of haloes with $M\lesssim10^{6}\Msun$, approximately the mass of the smallest dark matter dominated dwarf galaxy observed. WDM haloes also appear to have inner profiles that are less concentrated than their CDM counterparts \cite[][]{maccio2012,maccio2013}.

\par
However, satellite galaxies are not the only probes of the small-scale matter distribution. Gravitational lensing can be used to test whether dark matter is cold or warm. In particular, strong lensing, where a background galaxy is severely distorted by a foreground cluster, producing arcs and even Einstein rings, can probe the fine-scale matter distribution in a cluster \cite[e.g.][]{xu2009,meneghetti2011,killedar2012}. We show in a companion paper, \cite{mahdi2014a}, that there are clear observational differences in the lensing distribution between WDM \& CDM, {\em despite} the scales probed being well above the free-streaming scale. In this companion study, we find that, contrary to the naive expectation, WDM clusters have {\em higher} lensing efficiencies than CDM clusters {\em despite} containing less substructure. Here we explore why this would be the case by examining the mass accretion history and internal properties of WDM clusters. We show that the differences in the formation and growth of structure between WDM and CDM haloes leaves an imprint even at cluster scales, which are significantly greater than the free-streaming scale. The resulting substructure distribution leaves an distinct imprint on the gravitational lensing signature of cluster mass objects, which are at several orders of magnitude above the free-streaming scale in our WDM model. 

\par
This paper is organized as follows: we briefly describe the numerical methods in \Secref{sec:methods}, present our findings in \Secref{sec:results} and discuss these surprising results in \Secref{sec:discussion}.

\section{Numerical Methods}\label{sec:methods}
\begin{figure*}
    \centering
    \includegraphics[width=0.95\textwidth]{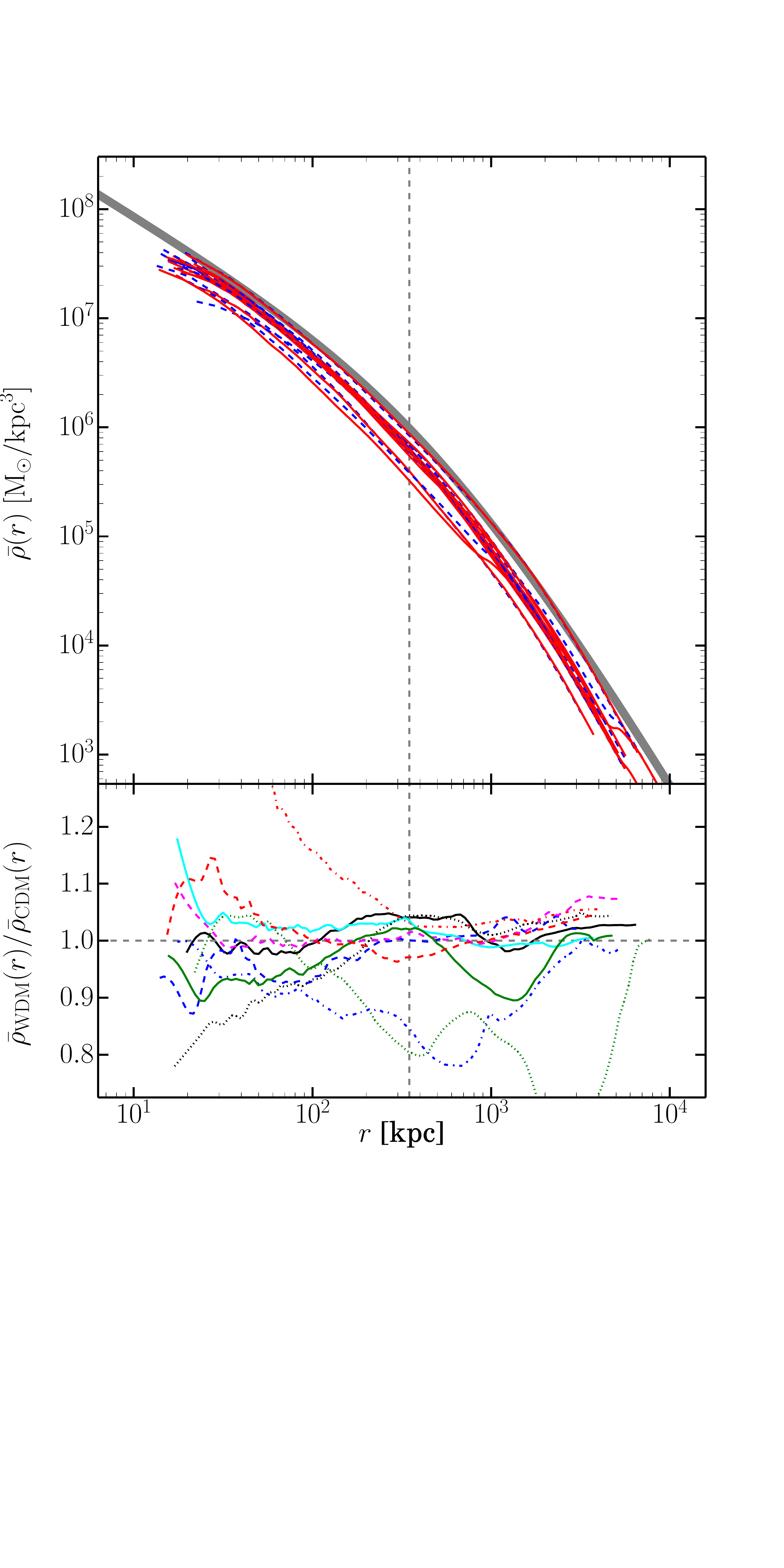}\vspace{-1pt}
    \includegraphics[width=0.95\textwidth]{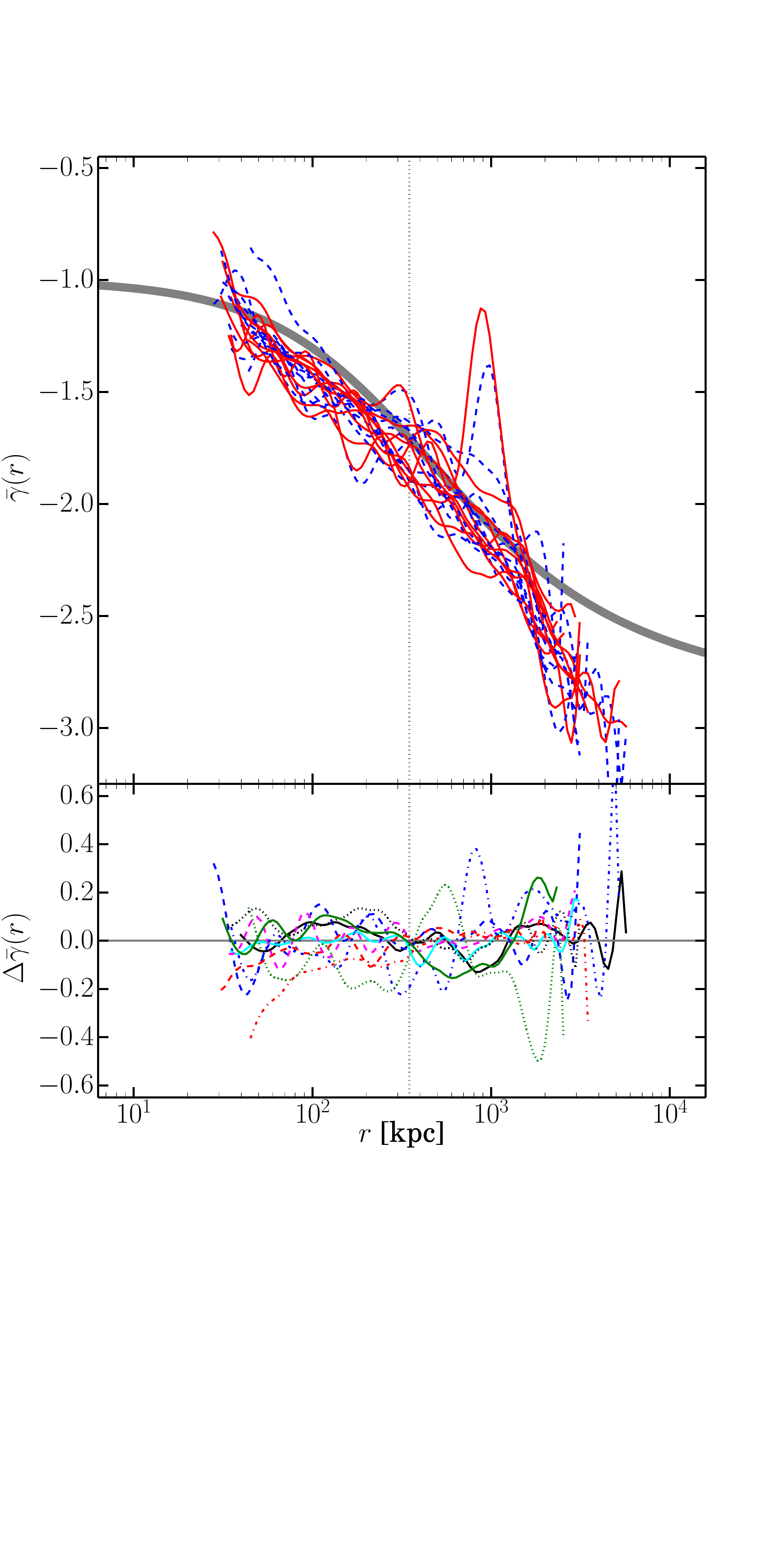}
    \caption{Smoothed projected density images of the high resolution region at $z=0$ centred on the largest cluster. Top panel shows CDM simulation and bottom panel shows its WDM counterpart. Dark regions are dense. The large cluster studied in this simulation is outlined by a solid blue circle. Also highlighted by a dashed red rectangle is a filament, which in the WDM simulation shows the artificial ``beads on a string'' effect, where a filament has fragmented into several spurious haloes.}
    \label{fig:sims}
\end{figure*}
We focus our study on 10 pairs of clusters extracted from zoom simulations of WDM and CDM cosmologies, which we summarize here. The zoom simulations used a parent simulation of $L_{\rm box}=150\Mpch$ containing $128^3$ particles with the following cosmological parameters: $h=0.7$, $\Omega_m=0.3$, $\Omega_{\Lambda}= 0.7$, and $\sigma_8=0.9$. Clusters with masses of $\geq10^{14}\Msunh$ were identified using \ahf\ ({\small{\textbf{A}MIGA}'s \small{\textbf{H}}alo \small{\textbf{F}}inder}; cf. \citealt{ahf}). For each selected cluster, all particles within a radius of $\sim 3 R_{\rm vir}$ of the cluster at $z=0$ are identified and their initial positions are altered using an inverse Zel'dovich transformation to obtain the Lagrangian positions at $z=\infty$. This initial Lagrangian volume is then populated with particles of the desired resolution, here $\sim1.4\times10^{9}\Msunh$, and these are perturbed using density perturbations of the parent simulation plus power from modes down to the new resolution scale. The resampled Lagrangian regions are initialized using $\Lambda$WDM \& \LCDM\ cosmologies, which primarily differ in the power included at small scales. The WDM model used is a $0.5$~keV thermally produced dark matter particle \cite[][]{bode2001,power2013}, which results in a suppression of growth for halo with $M\lesssim M_{\rm hm}=2.1\times10^{11}\Msun$, the so-called half-mode mass scale where the WDM power spectrum is 1/4 that of the CDM one \cite[][]{schneider2012}. We should note that the WDM initial conditions {\em do not} include non-gravitational velocities, thus technically, the WDM simulations are CDM ones with a smooth truncation in the initial power spectrum of density perturbations at a scale corresponding to $0.75\Mpch$\footnote{Adding in thermal velocities in the initial conditions can lead to the formation of  spurious low-mass haloes at late times, distinct from the problem noted in recent studies \cite[e.g.][]{wang2007},  see \cite{power2013}.}. All simulations were run with {\small GADGET2}, a TreePM code \citep{gadget2}. All simulations used the same time-stepping criteria. Each pair of zoom simulations used the same gravitational softening length, which was based on \cite{power2003}, ie: $\epsilon_{\rm opt}=4\,R_{\rm vir}/\sqrt{N_{\rm vir}}$ using $R_{\rm vir}$ from the parent CDM simulation. Each simulation produced 26 snapshots, spaced by $\Delta\ln a=1.0964$ from $z=9$, giving time intervals of $\sim200$~Myr at $z\gtrsim2$ to $\approx 1$~Gyr at low redshifts.

\par
For every snapshot of each zoom simulation we identify all bound haloes and their substructures using \velociraptor\ (formerly known as the {\bf ST}ructure {\bf F}inder, \stf) \citep{elahi2011}. This code identifies field haloes using a 3D Friends-of-Friends (FOF) algorithm with a linking length of $0.2$ times the inter-particle spacing \cite[see][for a description of FOF algorithms]{fof}. Substructures are identified by utilising the fact that dynamically distinct substructures in a halo will have a local velocity distribution that differs significantly from the mean, i.e. smooth background halo. The code first identifies particles that appear dynamically distinct from the halo background by examining the local velocity distribution and then links this outlier population using a phase-space FOF approach \cite[for details see][]{elahi2011}. The linking length criteria used for the phase-space FOF are the default used in previous studies \cite[e.g.][]{onions2013a,elahi2013a}, which will identify subhaloes along with the unbound tidal debris associated with subhaloes. We should mention that \velociraptor\ can also identify completely disrupted tidal debris \cite[][]{elahi2013a}, however here we are not interested in tidal streams. Therefore, we limit the search to subhaloes with tidal features by requiring that a particle has potential energy that is at least $0.5$ times that of its kinetic energy relative to the centre-of-mass of the substructure, eliminating completely unbound tidal debris in the unbinding procedure. 

\par
Running WDM simulations is not trivial as there is no agreed way of including the effects of the sub-resolution free-streaming motions of the warm dark matter particle in discrete N-body simulations. We show an example of the dark matter distribution around a cluster for one of our pairs of simulations in \Figref{fig:sims}. The overall large-scale structures are similar but the CDM cosmology contains many more dense small haloes than the WDM cosmology. We also highlight a filament in this figure to highlight a consequence of the absence of density perturbations at small scales. The region illustrates that WDM simulations are prone to artificial clumping of particles at scales below the half-mode mass scale \cite[e.g.][]{angulo2013,schneider2013a}. The dense clumps in the WDM filament are equally spaced apart and most do not have analogues in the CDM model. \cite{lovell2013} show that these spurious objects arise from highly elongated disc or needle like particle distributions in the initial Lagrangian particle distribution whereas genuine subhaloes arise from more spherical particle distributions. \cite{power2014a} argue that their formation arises because of discreteness effects during the initially anisotropic collapse of the density field. The result is that the mass function of ``haloes'' identified by a FOF halo finder will show up as a sharp upturn at a mass scale which depends on the resolution. These artificial haloes can survive accretion and appear as highly flatten density peaks in a halo. These spurious subhaloes are incorrectly identified as true subhaloes by density based subhalo finders such as {\sc subfind} \cite[][]{subfind}, typically showing up as another subhalo population at masses of $\lesssim\tfrac{1}{2}M_{\rm hm}$. \velociraptor\ uses dynamical information to identify subhaloes and may be slightly less prone to mistakenly identifying {\em subhaloes} arising from artificial clumping than density based subhalo finders. Conversely, {\em haloes} are identified using a simple 3D FOF algorithm like that used by {\sc subfind}, consequently the population of haloes composed of $\lesssim100$ particles will be dominated by these apparently spurious objects. Fortunately, this only applies to poorly resolved haloes and those composed of $\gtrsim100$ particles, and subhaloes originating from them, are not numerical noise. We discuss this issue in more detail in \Secref{sec:subhaloes}.

\par
To study the halo's mass accretion history, we use the halo merger tree algorithm associated with \velociraptor\ \cite[see][for a discussion of the code]{srisawat2013}. This algorithm is a particle correlator: that is the algorithm compares two (or more) exclusive particle ID lists and produces a catalogue of matches for each object in each catalogue. Specifically, for each object $i$ in catalogue $A$, the algorithm finds all objects $j$ in catalogue $B$ that share particles and calculates $\mathcal{M}_{ij}=N_{A\cap B}^2/(N_A N_B)$ to determine the merit of the initial matches. When constructing the halo merger tree, the match that maximises $\mathcal{M}$ both backward and forward in time is deemed the primary progenitor as haloes can only have one descendant. This code is also used to find the WDM counterparts to the CDM (sub)haloes. 

\par
For each (sub)halo, we calculate its virial mass, total mass, and maximum circular velocity $V^2=GM(r)/r$, the associated radius, mass, and density. Unless stated otherwise, the density is the average enclosed density, 
\begin{align}
  \bar\rho(r)&\equiv\frac{M(r)}{4\pi r^3/3}.
\end{align}
We also determine the logarithmic mass growth, 
\begin{align}
  \alpha_i\equiv\frac{d\ln M}{d\ln a}=\frac{\ln\left(M_i/M_{i-1}\right)}{\ln\left(a_i/a_{i-1}\right)},
  \label{eqn:massgrowth}
\end{align}
where $M$ and $a$ are the mass and scale-factor, respectively, at the $i^{\rm th}$ snapshot. We focus on the secular mass growth, defined as the mass change of the primary progenitor due to the accretion of mass {\em not} contained in previously identified haloes, that is excluding the mass change due to mergers. 

\par
Additionally, for the clusters we fit the average enclosed density profiles by commonly used \citetalias{nfw} \cite[][]{nfw} profile, 
\begin{align}
  &\rho(r)=\frac{\rho_{\rm NFW}}{(r/\rnfw)(1+r/\rnfw)^2}, \quad\text{where}\\
  &\bar\rho_{\rm NFW}(x\equiv r/\rnfw)=3\rho_{\rm NFW}\left[\frac{\ln(1+x)-x/(1+x)}{x^3}\right], \label{eqn:nfw}
\end{align}
We should mention that, although the \nfw\ profile is not a poor description of halo density profiles, many studies show that an Einasto profile
is a better fit to simulation data \cite[e.g.][]{gao2008,navarro2010}. Even the Einasto profile forces a particular functional form for the logarithmic slope $\bar\gamma(r)\equiv d\ln\bar\rho/d\ln r$. To avoid this somewhat artificial constraint and directly compare slopes in order to determine whether WDM haloes are more cuspy or cored than their CDM counterparts, we use a more generalised fit using B-splines. As B-splines are not commonly used in astronomy, we give a brief introduction here \cite[for more information see for instance][]{deboorpracticalguidetosplines1978,rogersintrotonurbs2001}. A B-spline function $f(x)$ is a linear combination of some constant coefficients $a_i$ with some polynomial functions $B_{i,k}(x)$ (B-spline basis functions) of a given degree $(k-1)$, i.e. $f(x)=\sum_i a_i B_{i,k}(x)$. These polynomial functions $B_{i,k}(x)$ are smooth and consist of polynomial pieces joined together in a special way. Fitting these basis functions to the data allows extra degrees of freedom which can give an unbiased fit to the underlying data when compared to a model with a fixed functional form. To determine $\bar\gamma$, we use a least squares 3$^{\rm rd}$ order B-spline fit to the enclosed density profile \footnote{Note that fitting B-splines requires a set of control points and one can also impose additional constraints such as a smoothing penalty in order to reduce the amount oscillations in the fit. For simplicity, we use a uniform distribution of points in $\ln r$ and do not impose any smoothing penalty on the second derivative of the density distribution.}.

\section{Results}\label{sec:results}
\subsection{WDM clusters}\label{sec:clusters}
\begin{figure}
    \centering
    \includegraphics[width=0.875\columnwidth, trim=0cm 16.cm 2.5cm 5cm, clip=true]{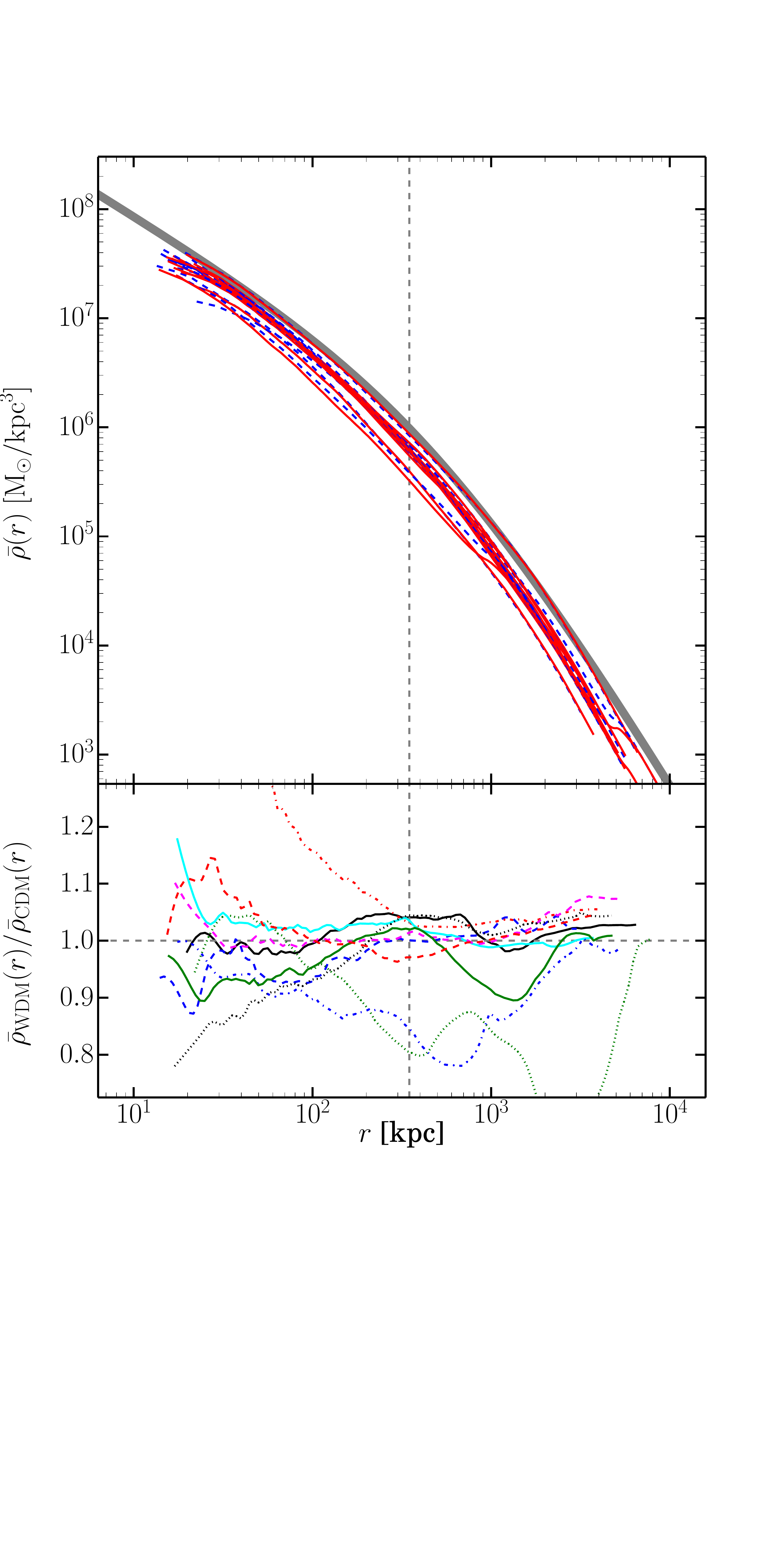}
    \includegraphics[width=0.875\columnwidth, trim=0cm 14cm 2.5cm 5.25cm, clip=true]{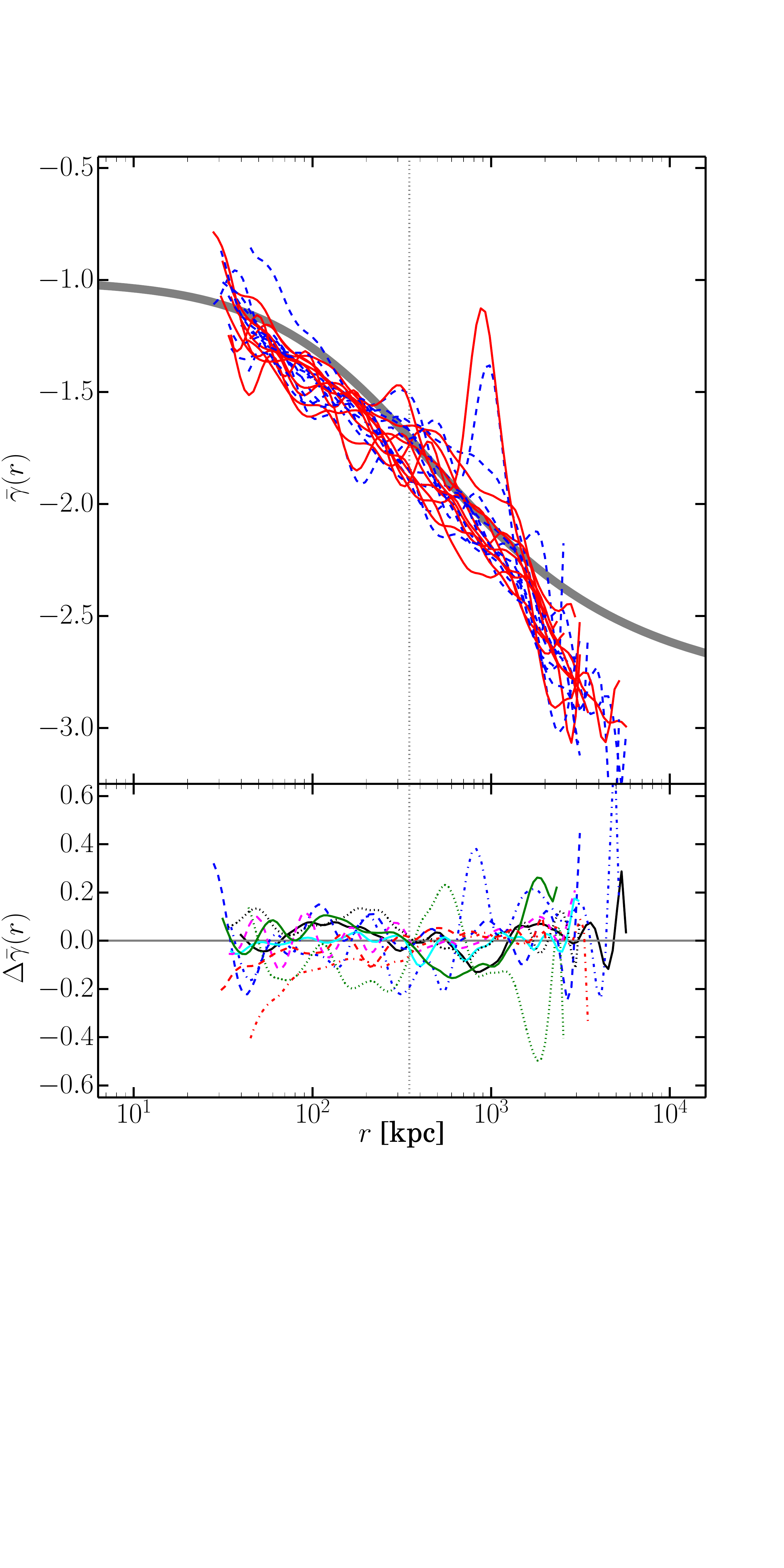}
    \caption{Average enclosed density $\bar\rho(r)$ for WDM \& CDM clusters plotted as solid red and dashed blue lines respectively. Top two panels show the density and the ratio of WDM to its CDM counterpart, each cluster with a different line type and colour. The bottom two panels show the logarithmic slope along with the difference between WDM clusters and their CDM counterparts. We also plot thick grey lines corresponding to an \nfw\ fit of the most massive, densest cluster and a vertical dashed line at the scale radius $\rnfw$. The density ratio panel \& the $\Delta\bar\gamma$ plot have horizontal lines at $1.0$ and $0$ respectively to guide the eye.}
    \label{fig:denprofiles}
\end{figure}
The removal of small-scale power not only suppresses the growth of structure below the WDM dampening scale as seen in \Figref{fig:sims}, here for haloes with masses of $\lesssim2\times10^{11}\Msun$ or for haloes composed of $\lesssim100$ particles, it also alters the internal properties of clusters, which are 3 orders of magnitude larger than $M_{\rm hm}$. The lack of small-scale power has a small but noticeable effect on the density profiles of haloes as shown in \Figref{fig:denprofiles}. Here we plot the $z=0$ enclosed average density profiles of our ten clusters with the substructures {\em excluded}. For reference, we also show a least squares \nfw\ fit to the most massive, densest cluster by a thick grey line. First, we note that the WDM \& CDM profiles differ but typically only by $\sim10\%$ as shown in the second panel in \Figref{fig:denprofiles}. Three clusters have significantly different profiles: one in which the WDM counterpart is significantly denser for much of the central region; and two others where the reverse is true. The \nfw\ profile is a reasonable approximation to the density, although for the reference curve shown, it overestimates the density in the very centre and in the outermost regions. In the example shown, the \nfw\ fit using a nonlinear Levenberg-Marquardt least squares gives $\rnfw\approx345\pm5$~kpc for both the WDM cluster and its CDM counterpart. Though most clusters have similar scale radii, the three clusters with different profiles have fitted $\rnfw$ that disagree at the 4$\sigma$ level. 

\par
The logarithmic slope, $\bar\gamma$ calculated by fitting B-splines to the density profile is shown in the lower two panels of \Figref{fig:denprofiles}. WDM clusters have similar $\bar\gamma$ to their CDM counterparts and there appears to be no significant trend to more cuspy or cored density profiles at cluster mass scales. Again, here we show the slope of the \nfw\ fit to the largest cluster for reference. Typically, the slope inferred by fitting B-splines is steeper that the corresponding \nfw\ profile in the very inner and very outer regions. The local oscillations in the logarithmic slope are {\em not} significant and can be damped by choosing a smoothing penalty when fitting a density profile. Here for simplicity we do not impose such a penalty. Note that the cluster with a significant change in $\bar\gamma$ at $\approx1$~Mpc is one in close proximity to another large cluster and is just beginning to merge with this neighbour, resulting in a deviation of the density in the very outer regions, present in {\em both} WDM \& CDM. 

\begin{figure}
    \centering
    \includegraphics[width=0.875\columnwidth, trim=0cm 14cm 2.5cm 5.25cm, clip=true]{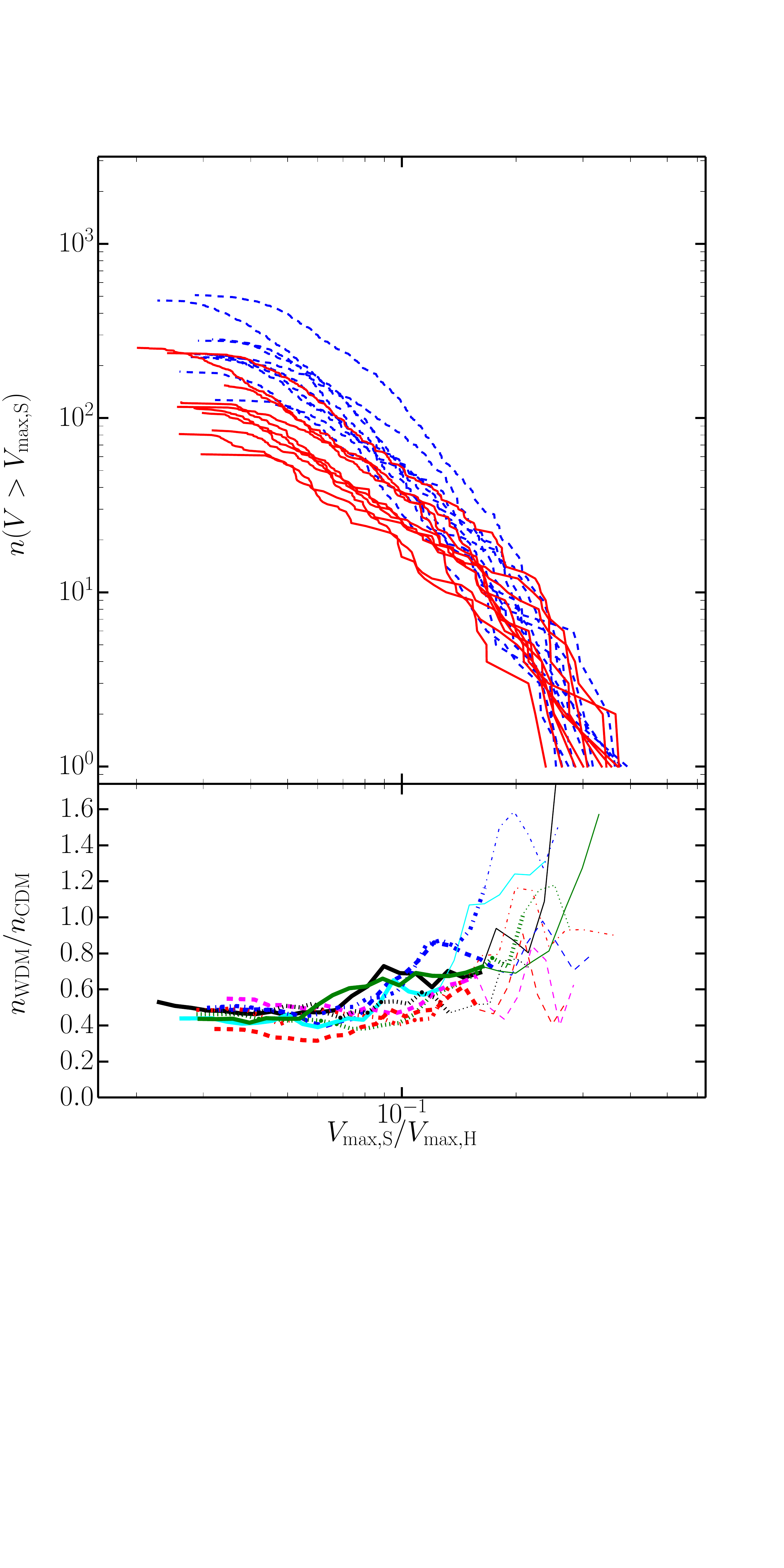}
    \caption{Cumulative maximum circular velocity distribution of subhaloes for each cluster. Top panel is the cumulative distribution function, bottom panel is the ratio. Line styles are the same as in \Figref{fig:denprofiles}. In the bottom panel, we also emphasise the region where $n(V>V_{\rm max})>10$, that is the region where uncertainties are $\lesssim10\%$, by thick lines.}
    \label{fig:subvmaxcdf}
\end{figure}
\par
From this one might conclude that WDM \& CDM clusters, which sample scales that are $\sim10^3M_{\rm hm}$, are similar. This is not the case as their internal properties do show significant differences, as visually seen in \Figref{fig:sims}. From the substructure distribution, shown in \Figref{fig:subvmaxcdf}, it is quite evident that WDM clusters contain fewer substructures than their CDM counterparts. From the lower panel, which shows ratio of the cumulative distribution functions (CDF), we see that at low $V_{\rm max}$, the ratio does not appear to have a pronounced slope, indicating that the functional form, at least at the scales sampled here, does not change drastically. We do expect the CDF to turnover in the WDM case at smaller $V_{\rm max}$ that lie below our resolution limit of $\sim5\times10^{10}\Msun=0.5M_{\rm hm}$ as at smaller scales the progenitor haloes simply do not form \cite[see for instance][where the halo mass function is strongly suppressed for $M\lesssim0.1M_{\rm hm}$]{schneider2012,schneider2013a}. This figure also shows that WDM haloes are only slightly more likely to have subhaloes with high maximum circular velocities but in general the subhalo distribution of WDM cluster for these scales is simply a suppressed version of the CDM one. Finally, we should note that the lack of a significant upturn in the WDM subhalo velocity distribution indicates that artificial subhaloes are {\em not} a significant population at the scales of interest here \cite[see][for an example of the spurious subhalo velocity distribution]{lovell2013}. 

\par
\begin{figure}
    \centering
    \includegraphics[width=0.99\columnwidth]{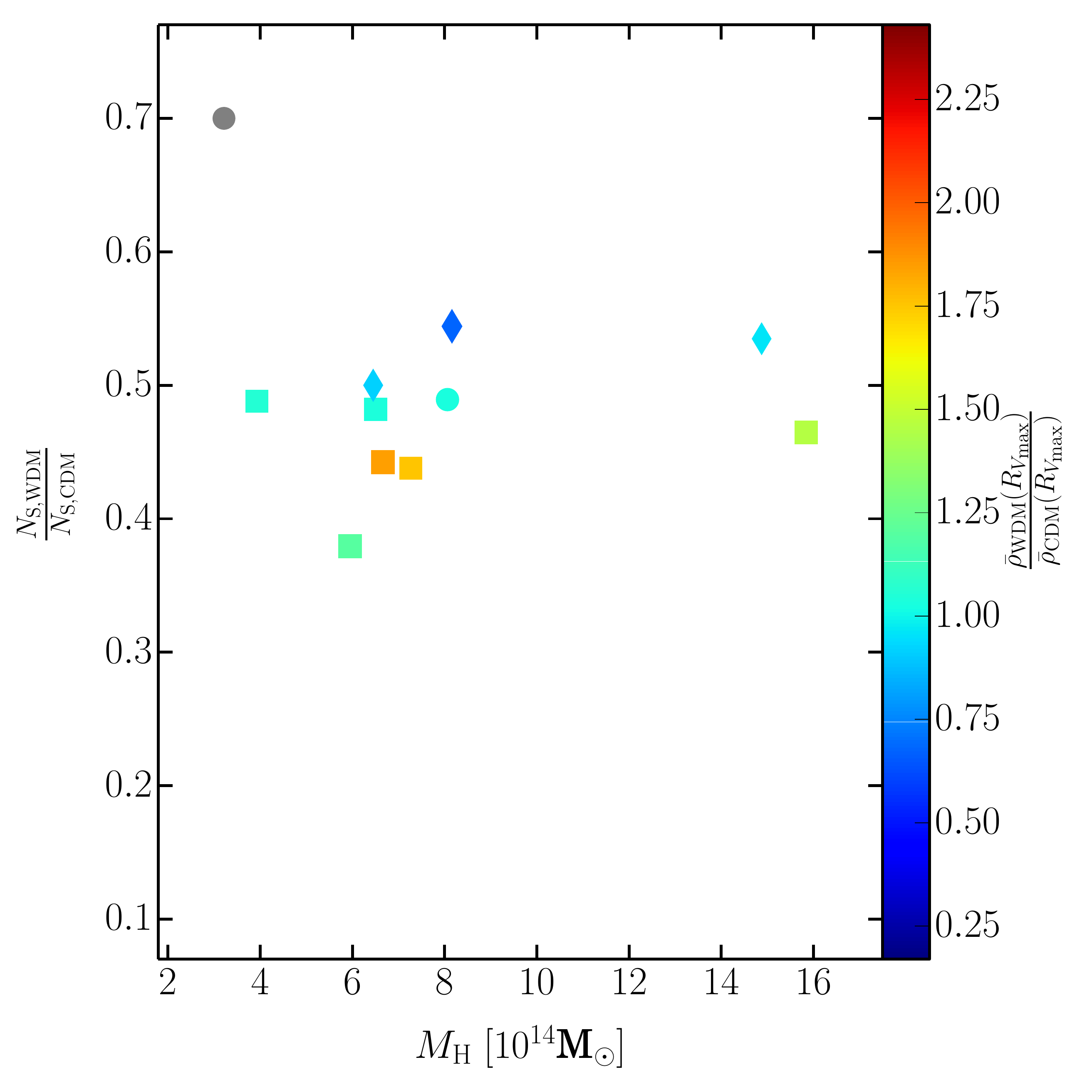}
    \caption{Comparison of the internal properties of WDM clusters to their CDM counterparts: we show the ratio of the number of substructures as a function of mass of the CDM cluster. Colour indicates the ratio of the average enclosed density within the maximum circular velocity radius and marker sizes scale with ratio of the masses. We show representative grey point to indicate the size of a halo with a mass ratio of 1. The points scale logarithmically with $M_{\rm H,WDM}/M_{\rm H,CDM}$. To emphasise the differences between WDM and CDM, we plot clusters where the WDM cluster has {\em higher} average central density and {\em larger} $\rvmax$ than its CDM counterpart as circles, those with {\em only higher} $\rho(\rvmax)$ as squares, and those with {\em only larger} $\rvmax$ as diamonds.}
    \label{fig:nsubcomp}
\end{figure}

We summarise the differences between WDM \& CDM clusters in \Figref{fig:nsubcomp}, specifically the total number of substructures, $N_{\rm S}$ and average densities within the radius of maximum circular velocity, $\bar{\rho}(\rvmax)\equiv M(R<\rvmax)/(4\pi \rvmax^3/3)$. WDM clusters have roughly half as many substructures as their CDM counterparts, down to $0.5M_{\rm hm}$. There is no correlation between the host halo's mass and $N_{\rm S, WDM}/N_{\rm S, CDM}$, in part due to the mass resolution of our simulations. We expect $N_{\rm S,WDM}/N_{\rm S,CDM}$ to decrease with decreasing $M_{\rm H}$ if we were able to resolve scales well below $M_{\rm hm}$. The colour of the points indicate the ratio of the central densities. Based on \Figref{fig:denprofiles}, we expect about half of the WDM clusters to have slightly higher central densities. This figures shows that most clusters have slightly higher average central densities in part due to having smaller $\rvmax$. The differences in $\rvmax$ are typically $\lesssim20\%$. Finally, the size of the points scale with $\log\left(M_{\rm H,WDM}/M_{\rm H, CDM}\right)$. The fact that all the points are visually the same size as the reference point, where this ratio is 1, indicate the masses of WDM clusters differ by only a few percent relative to the CDM counterparts.
\begin{figure}
    \centering
    \includegraphics[width=0.99\columnwidth]{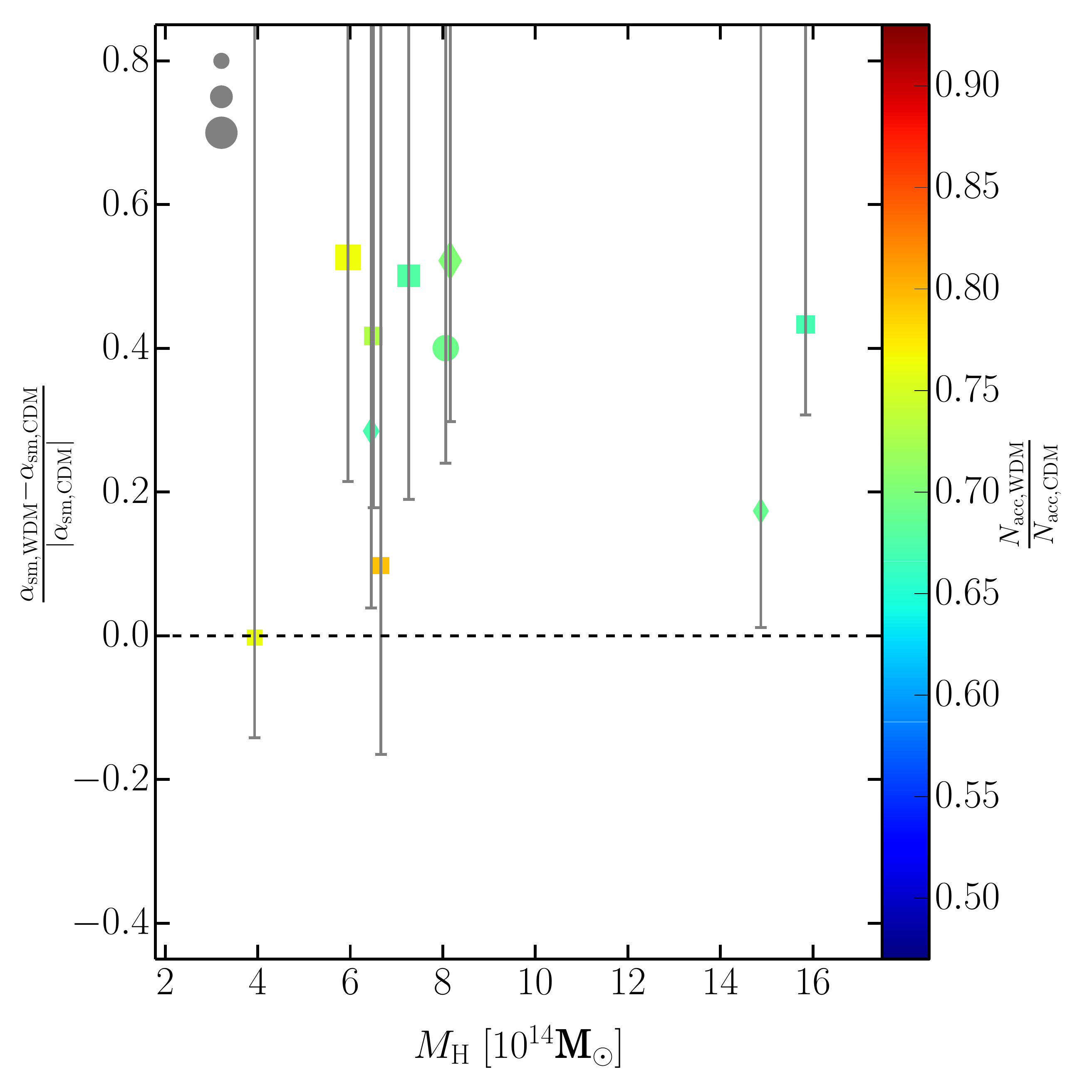}
    \caption{Comparison of the secular mass accretion history of haloes. Here we show the median ratio of the logarithmic secular mass growth between WDM clusters and their CDM counterparts over their lifetime, along with error bars enclosing a 1-$\sigma$ quantile ($34\%-66\%$). Colour indicates the ratio of the number of accretion events experienced by the main progenitor halo over its lifetime. Marker sizes scale linearly with the ratio of the number of merger events $N_{\rm merge,WDM}/N_{\rm merger,CDM}$. We show 3 representative grey points to indicate the size of a halo ratios of 0.5, 1, and 2 going from top to bottom. Marker styles are the same as in \Figref{fig:nsubcomp}.}
    \label{fig:secmasshistorycomp}
\end{figure}

\par
The $z=0$ differences are a result of differing mass accretion histories shown in \Figref{fig:secmasshistorycomp}. Here we have calculated the logarithmic mass growth in the WDM case and compared it to the CDM case for every single consecutive pair of snapshots, i.e., $\left(\alpha_{\rm WDM}-\alpha_{\rm,CDM}\right)/\left|\alpha_{\rm CDM}\right|$. We then show the median of this distribution for the entirety of the halo's history along with a $1-\sigma$ quantile by an error bar. In \Figref{fig:secmasshistorycomp}, we focus on {\em smooth} mass accretion, i.e. accreted mass that was not identified as belonging to a FOF halo composed of at least 20 particles in the previous snapshot. We should note that as a result of the temporal spacing used, $\Delta\ln a=1.0964$, $\alpha$ is a biased estimator, and represents an upper limit to the smooth mass accretion rate. With shorter times between snapshots, we would be able to track more haloes, those that formed in between the snapshots. Though both cosmologies are equally affected by the temporal bias in $\alpha$, the presence of spurious haloes in WDM simulations artificially biases this estimate to lower values. Fortunately, the artificial bias in WDM models is mitigated by the fact that these spurious objects, though large in number, account for a small fraction of the mass increase experienced by the main cluster halo.

\par
Despite this biased estimator, if the removal of small scale power did not systematically affect the form of mass accretion, one would expect the distribution to be centred along zero (indicated by the dashed horizontal line). This is evidently not the case. The range in the ratio is very broad and there are stages where the CDM host undergoes higher smooth mass accretion rates than their WDM counterparts, but these epochs are not common. Even at cluster scales, WDM haloes accrete mass from the surrounding environment at a rate that this typically $\sim40\%$ higher than their CDM counterparts. Taking into account the presence of spurious haloes in WDM simulations, the smooth accretion rate is {\em even higher}. Correspondingly, WDM haloes have significantly fewer accretion events, i.e., where the host halo accretes a smaller halo. However, this trend is not present if one is looking at the number of {\em major mergers} a halo has undergone, $N_{\rm merge}$. Here we define a merger as an accretion events where the mass ratio between the main progenitor and a secondary is $\geq1/3$. The data points scale linearly with ratio of WDM mergers to CDM mergers and for reference we have drawn 3 points showing the size for ratios of $0.5,1$ and $2$. The variety of marker sizes clearly indicates that major mergers are not systematically affected by the removal of small-scale power.

\par
Thus at cluster scales, the higher smooth accretion rate and absence of small scale structure in a WDM cosmology has slightly increased the average enclosed density within the maximum circular velocity radius, significantly reduced the number of subhaloes but generally left the densities profiles unchanged. Smaller scales should show even greater differences.

\subsection{WDM Subhaloes}\label{sec:subhaloes}
\begin{figure}
    \centering
    \includegraphics[width=0.80\columnwidth]{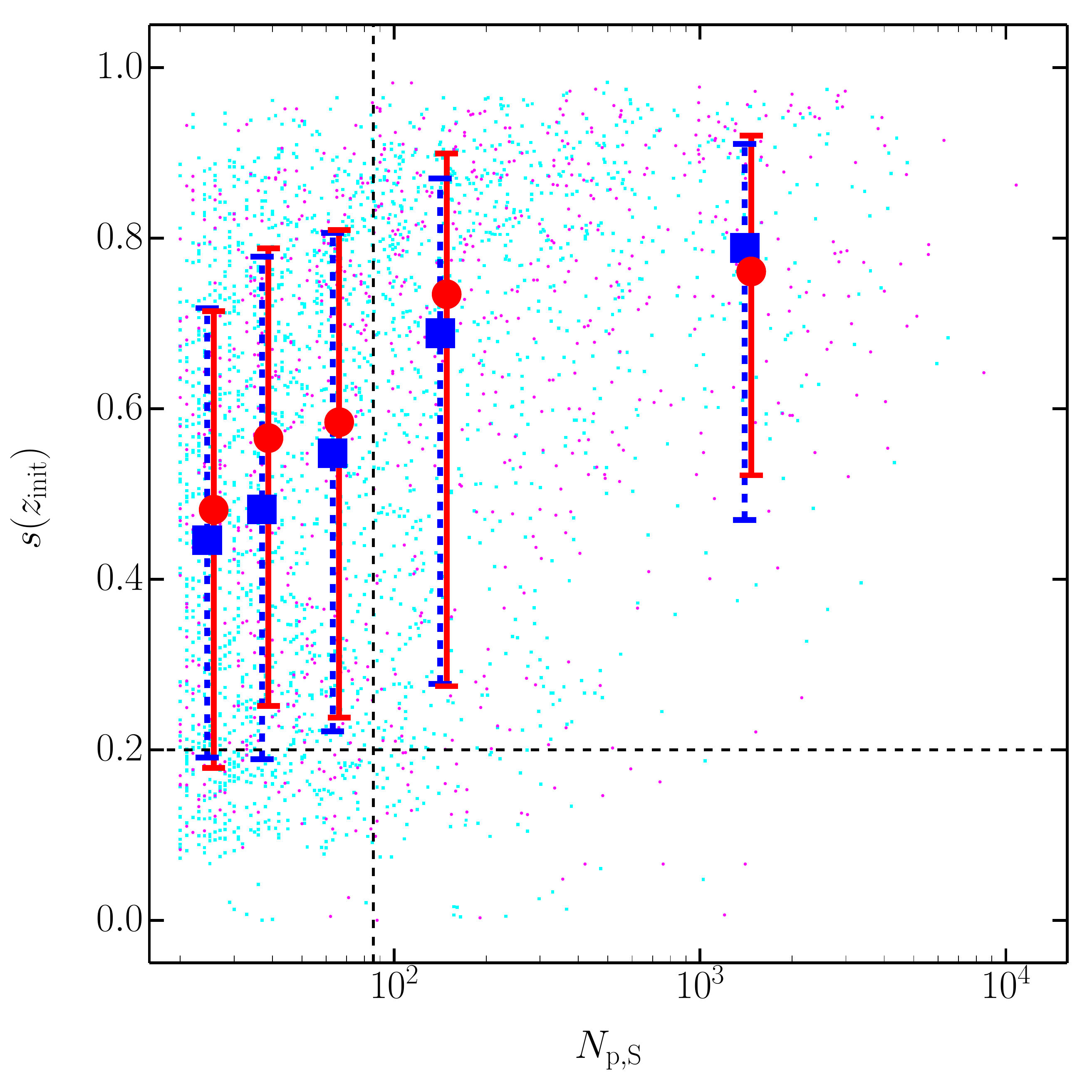}
    \caption{Minor axis of initial particle distribution for subhaloes identified at $z=0$ versus current particle number. WDM \& CDM subhaloes are shown in magenta \& cyan respectively. We also show the median and $16\%$ and $84\%$ quantiles in five different particle number bins, WDM in red and CDM in blue where mass bins are chosen to ensure equal number of {\em CDM subhaloes per bin}. For clarity we offset the WDM points by a small amount. We also show a horizontal dashed line corresponding to $s=0.2$, the cutoff in sphericity used by Lovell et al. (2014) to remove spurious WDM (sub)haloes and a vertical dashed lined at the particle number corresponding to $M_{\rm hm}$.}
    \label{fig:minoraxishighz}
\end{figure}
\begin{figure}
    \centering
    \includegraphics[width=0.80\columnwidth]{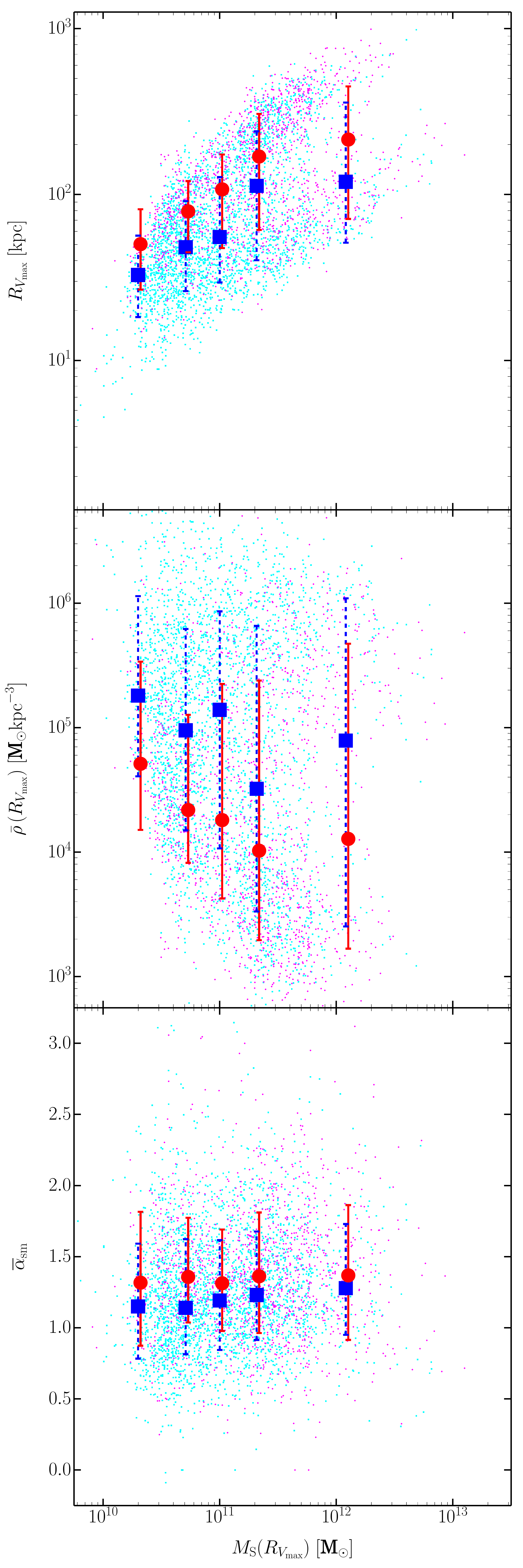}
    \caption{Properties of all substructures identified in the ten clusters studied at $z=0$. Top: the radius of maximum circular velocity, $\rvmax$. Middle: the enclosed average density $\bar{\rho}(\rvmax)$. Bottom: the median smooth logarithmic mass accretion rate of the progenitor halo over the lifetime of the progenitor halo. All are plotted as a function of mass within $\rvmax$. Style is similar to \Figref{fig:minoraxishighz}. 
   }
    \label{fig:subhaloprops}
\end{figure}
Before we discuss the properties of subhaloes in detail, it is important to note measuring the (sub)halo distribution in WDM simulations is not trivial. As mentioned in \Secref{sec:methods}, WDM simulations are contaminated by spurious objects below $M_{\rm hm}$ for haloes composed of $\lesssim100$ particles. As a consequence, both the halo and subhalo distribution as measured by a configuration space based halo finder, such as FOF or {\sc subfind}, are composed of two populations, real (sub)haloes and spurious ones. \cite{lovell2013} found that spurious haloes that comprise this secondary population, which appears at mass scales below $M_{\rm hm}$, originate from highly elongated distributions in the initial Lagrangian particle positions. They found that a distinct change in the median minor to major axis ratio of the reduced inertia tensor of the initial conditions in WDM models below $M_{\rm hm}$ relative to CDM models, peaking at minor to major axis ratios of $\approx0.1$ as compared to $\approx0.4$ in CDM haloes \cite[see also discussion in][]{power2014a}. Their study probed much smaller scales relative to $M_{\rm hm}$ than we do here, but the difference seen between WDM and CDM high $z$ morphology should still be seen in our simulations. If we examine the Lagrangian distribution of the progenitor haloes of the subhaloes identified by \velociraptor, which is a phase-space based finder that uses criteria based on a halo's internal dynamics, we find that there is {\em no significant} difference in the morphology of particle distribution from which subhaloes identified at $z=0$ formed. Specifically we show in \Figref{fig:minoraxishighz}, the ratio of the minor to major axis ratio from the reduced inertia tensor, $s$, of the initial particle distribution of the progenitor halo from which a subhalo now composed of $N_p$ particles formed. Unlike \cite{lovell2013}, the WDM subhalo population is not offset from the CDM one nor is $s$ small. Based on this figure and the lack of a significant second population with a different cumulative velocity distribution in \Figref{fig:subvmaxcdf}, we are confident that we are not dominated by spurious subhaloes, particularly at the mass scales of interest, $\gtrsim100$ particles.

\par
The properties of subhaloes of these clusters, which probe these smaller mass scales ($M\lesssim10^{13}\Msunh$ or $\lesssim100M_{\rm hm}$), is shown in \Figref{fig:subhaloprops}.  Here we show the size, central densities and smooth accretion rates, where we have chosen $\rvmax$ to define the scale of subhaloes as this region is less prone to tidal disruption \footnote{For reference, this radius occurs at $2.1626\times r_{\rm NFW}$ the characteristic scale for a \nfw\ halo}. We also plot the median, upper and lower quantiles for several mass bins, chosen such that there are equal number of CDM subhaloes in each bin. The median and quantiles show that WDM subhaloes are on average twice as large, but are only a factor of $\sim5$ times less dense. Naturally, the size depends on the mass of the object, however, the enclosed density does not appear to have a strong mass dependence. The fact that the WDM cluster mass haloes have slightly higher central densities and slightly smaller sizes (see \Figref{fig:nsubcomp}) suggests that the differences may be a result of both the mass scales probed and small number statistics. 
\begin{figure*}
    \centering
    \includegraphics[width=0.99\textwidth,trim=0cm 0cm 0cm 0cm,clip=true]{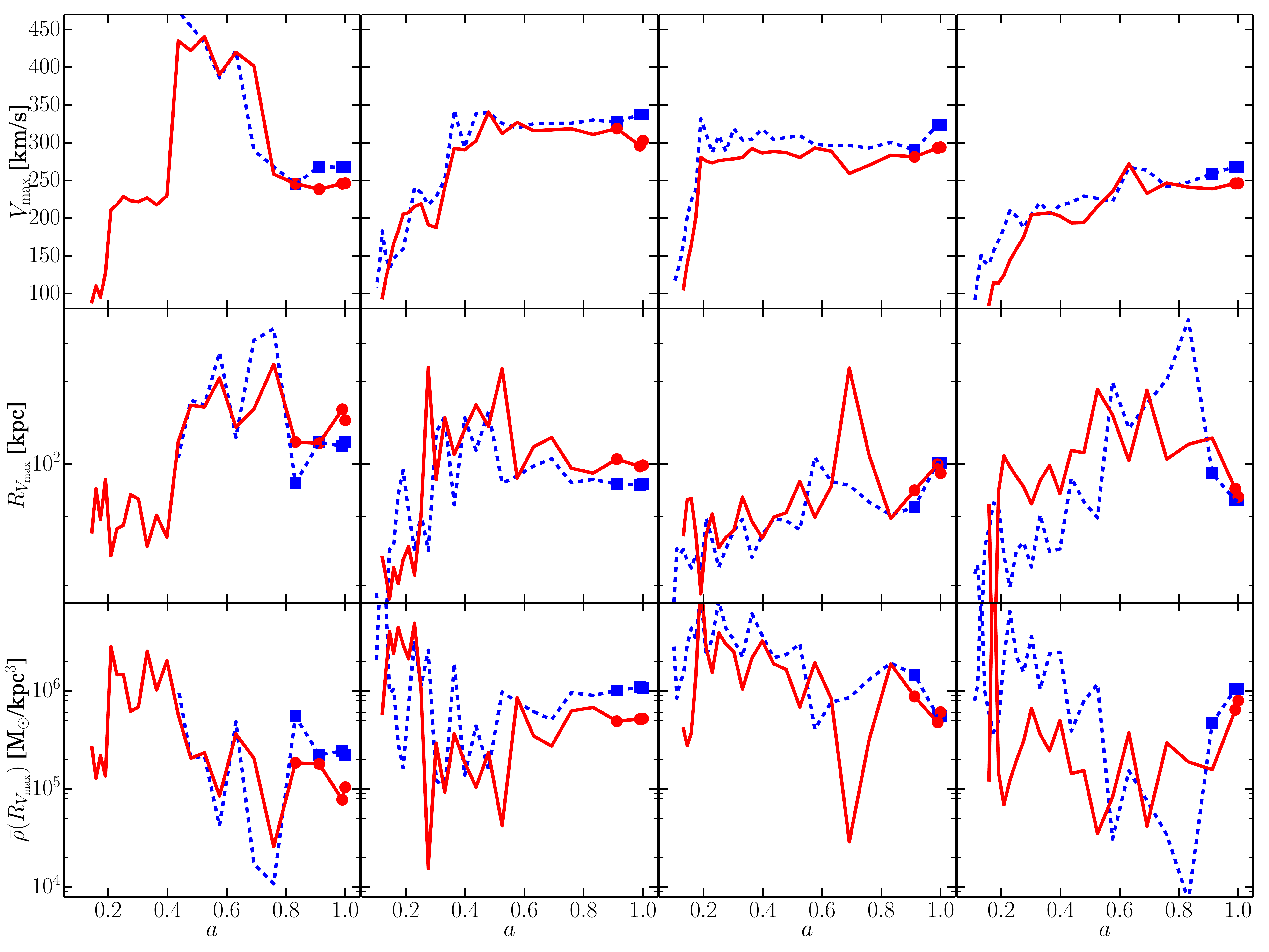}
    \caption{The evolution of 4 WDM subhaloes (red solid lines) and their CDM counterparts (blue dashed lines). Panels show $\vmax$ (top), $\rvmax$ (middle) and average density within $\rvmax$ (bottom). The snapshots where the object is a subhalo are marked by a filled circle and square for WDM and CDM respectively. Note that the CDM progenitor halo in the left-most column forms later than its WDM counterpart.}
    \label{fig:subevo}
\end{figure*}

\par
The differences between WDM haloes and CDM (sub)haloes is partially a consequence of the progenitor haloes having accreted more of their mass via smooth accretion as seen in the bottom panel of \Figref{fig:subhaloprops}. WDM accrete at rates that are $\sim0.15$ higher regardless of mass. Consequently, WDM haloes should have $\sim10-20\%$ more of their mass smoothly deposit in outer regions of the halo rather than having this mass in discrete overdensities, which can be drawn into the centre via dynamical friction.

\par
We show the evolutionary history of several large WDM subhaloes identified in the most massive cluster at $z=0$ along with their CDM counterparts in \Figref{fig:subevo}, specifically evolution of $V_{\rm max}$, $\rvmax$ and the average central density, where the CDM counterparts were identified using the particle correlator used to build the merger tree. Note that almost every single large WDM subhalo composed of $\gtrsim1000$ particles has a CDM (sub)halo counterpart. At early times, the progenitor halo is poorly resolved and thus $\rvmax$ is very noise despite $\vmax$ showing smooth evolution. After $a\gtrsim0.3$, (sub)haloes have similar histories in both cosmologies. The slight differences in small-scale matter distribution and mass growth result in CDM progenitor halos with larger $\vmax$ than their WDM counterparts prior to accretion. There are periods where CDM progenitor haloes also have larger $\rvmax$ than the WDM analogue and visa-versa, with a slight bias towards WDM haloes being larger on average. Once accreted, the nonlinear environment does not significantly change $\vmax$ though $\rvmax$ typically decreases. The end result is for these 4 massive subhaloes the WDM ones are $5-30\%$ more extended and $10-50\%$ less dense than their CDM counterparts.

\par
\begin{figure}
    \centering
    \includegraphics[width=0.99\columnwidth]{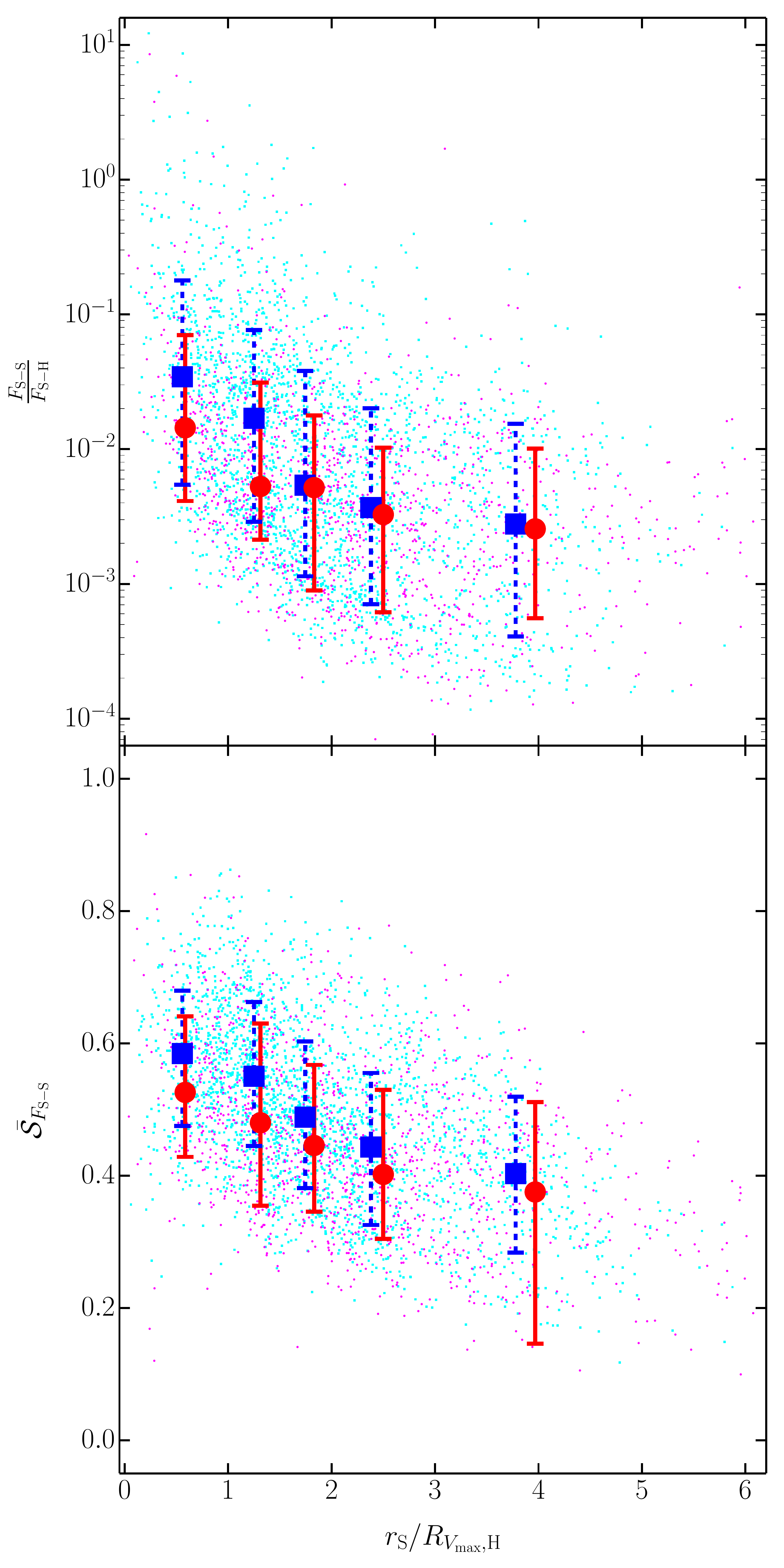}
    \caption{Gravitational force experienced by subhaloes. Top panel shows the ratio of the amplitude of the force arising from subhaloes to the force from the host as a function of radius. Bottom panel shows the product of the eigenvalues of the force distribution, a measure of how isotropic is the force arising from subhaloes. Error bars, markers, colours and line styles are the same as in \Figref{fig:subhaloprops}.}
    \label{fig:subhostforce}
\end{figure}

The evolutionary history suggests that the tidal disruption rate might not differ significantly between CDM and WDM cosmologies at these mass scales despite the fact that WDM haloes are more physically extended and not as dense, which should make them more prone to tidal disruption. Differences in the tidal disruption rate in part arise from differences the host density profile and subhalo-subhalo interactions. We show the ratio of the force experienced by a subhalo due to all other subhaloes relative to the force due to the host in the top panel of \Figref{fig:subhostforce} as a function of radial distance from the host. For simplicity we have treated subhaloes as point masses and assumed spherical symmetry, i.e.~:
\begin{align}
  {\bf F}_{{\rm S-S},i}=-\sum_{j\neq i}^{N_{\rm sub}}\frac{GM_{{\rm S}i}M_{{\rm S}j}}{|{\bf x}_j-{\bf x_i}|^3}({\bf x}_j-{\bf x}_i),
\end{align}
is the force on subhalo $i$ at position ${\bf x}_i$ relative to the halo centre due to all other $N_{\rm sub}$ subhaloes and 
\begin{align}
  {\bf F}_{{\rm S-H},i}=-\frac{GM_{{\rm S}i}M_{\rm H}(r_{i})}{r_{i}^3}{\bf x}_{{\rm S},i},
\end{align}
is the force on subhalo $i$ due to the host. The lower panel of \Figref{fig:subhostforce} shows how isotropic the force experienced by a subhalo due to all other subhaloes is via
\begin{align}
  \bar{\mathcal{S}}_{F_{\rm sub-sub}}=(\lambda_1\lambda_2\lambda_3)^{1/3},
\end{align}
where $\lambda_i$ are the normalised ordered eigenvalues of the force distribution
\begin{align}
  \boldsymbol{\mathcal{F}}_{i,(a,b)}=\sum_{j=1}^{N_{\rm sub}} \frac{M_{\rm S,j}}{|{\bf x}_j-{\bf x_i}|} \frac{({\bf x}_j-{\bf x}_i)_{a}({\bf x}_j-{\bf x}_i)_{b}}{|{\bf x}_j-{\bf x_i}|^2}.
\end{align}
An isotropic distribution gives $\bar{\mathcal{S}}=1$, whereas $\bar{\mathcal{S}}\ll1$ for highly anisotropic distribution.

\par
Figure \ref{fig:subhostforce} shows that the force due to subhaloes is small and is no longer negligible for radial distances of $\lesssim2\rvmax$. The ratio gradually increases with decreasing radius due to the higher subhalo density. CDM subhaloes typically experience a stronger subhalo force primarily due to the greater number of subhaloes in CDM cosmologies. If the force due to other subhaloes highly anisotropic, this would typically only alter the orbit of subhaloes whereas a more isotropic distribution is more likely to increase the rate of tidal stripping, by stirring the loosely bound outer regions of a subhalo. Again, only in the very central regions are there notable differences in the median value of $\bar{\mathcal{S}}$ between the two cosmologies, with CDM haloes experiencing a more isotropic pull than their WDM counterparts. The susceptibility of WDM subhaloes to tidal disruption is somewhat mitigated by the less harsh environment they experience. 

\par
\begin{figure}
    \centering
    \includegraphics[width=0.99\columnwidth]{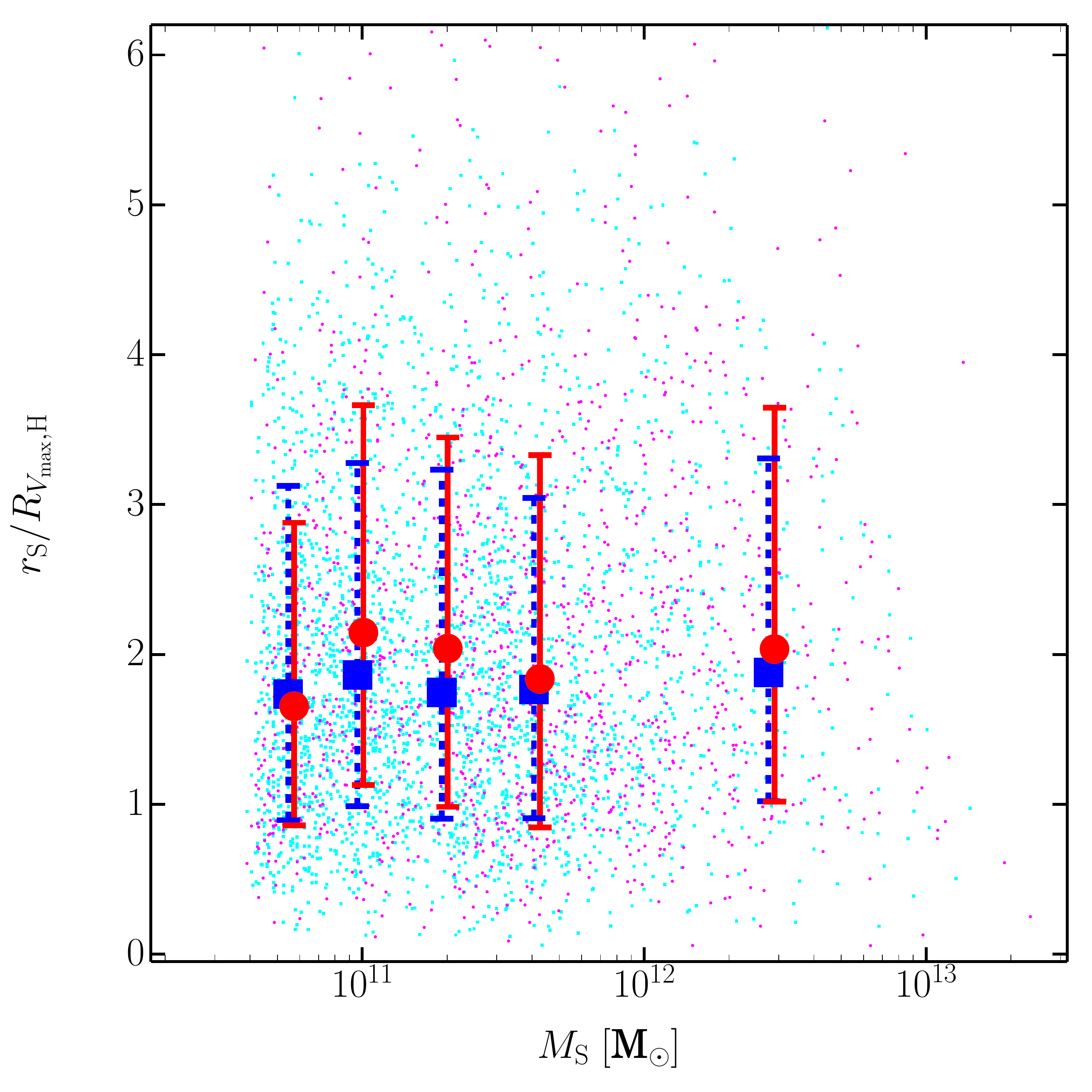}
    \caption{Distribution of the radial distance of a subhalo to the host halo's centre of mass scaled by $\rvmax$ of the host halo as a function of subhalo mass. Error bars, markers, colours and line styles are the same as in \Figref{fig:subhaloprops}.}
    \label{fig:subhalorad}
\end{figure}

Given subhalo-subhalo interaction are generally negligible for both cosmologies, the host halo should be the dominat force governing the radial distribution of subhaloes about the host. Since the host halo density profiles in CDM and WDM cosmologies are similar (see discussion in \Secref{sec:clusters}), one might predict the same to be true for the radial subhalo distribution with a slightly bias to larger radii for WDM subhaloes given they are more prone to tidal disruption. This expectation is illustrated in \Figref{fig:subhalorad}, which shows the median and quantiles of the radial distribution as a function of subhalo mass. The radial distributions overlap with WDM subhaloes have slightly higher medians and upper quantiles.

\par
The picture is then that WDM haloes relative to CDM haloes represent density distributions that contain fewer local density peaks, aka subhaloes. These secondary peaks are physically more extended with a lower average density and reside at greater distances from the central density peak. The question arises ``how can we distinguish between WDM and CDM clusters?''. Naturally the first answer that comes to mind is the galaxy distribution. Since WDM clusters have far fewer small satellites than their CDM counterparts, the galaxy luminosity function will differ significantly. The absence of satellites need not be the only signature of warm dark matter as analytic calculations by \cite{smith2011} using the halo model show that haloes are more biased in WDM models, a large-scale signature. Moreover, the mass scales at which there is a pronounced difference in the satellite population is at scales where feedback physics can drastically changed the baryonic content of a subhalo host. The uncertainty in the physics governing galaxy formation leads us to use another probe of the underlying matter distribution, namely gravitational lensing. We fully discuss the differences in the strong lensing distribution in our companion paper, \cite{mahdi2014a}. Here we simply present how the differences might manifest themselves in a lensing observation. 

\subsection{Seeing the Universe through a Warm Lens}\label{sec:lenses}
Under the thin lens approximation, the deflection angle produced by a gravitational lens is proportional to the lens's surface density. Strong lensing occurs when the mass surface density is close to or exceeds a critical value,  $\Sigma(r)\gtrsim\Sigma_{\rm crit}$, where
\begin{align}
  \Sigma_{\rm crit}\equiv \frac{c^2}{4\pi G}\frac{D_{\rm s}}{D_{\rm l}D_{\rm ls}},\label{eqn:sigmacrit}
\end{align}
where $c$ is the speed of light, $G$ is the gravitational constant and $D$ is the angular diameter distance to the source $s$, lens $l$ and between the lens and source $ls$ respectively. In fact, strong lensing can occur even in regions where $\Sigma(r)=\beta\Sigma_{\rm crit}$, with $\beta$ on the order of unity. Therefore, denser clusters will result in stronger lensing. Moreover, clusters with more substructure, that is local density peaks, should increase the area of a cluster that lies above the critical threshold. To see if the differences in the density profiles of subhaloes affect the lensing distribution we place our lens at $z=0.3$ and our source at $z=2$, and determine the radius at which $\bar\Sigma(r)=\beta\Sigma_{\rm crit}$, where we set $\beta=0.2$. We assume that each (sub)halo is a spherical density described by an \nfw\ profile with $\rnfw=a\rvmax$ and treat it in isolation. We then numerically solve for when 
\begin{align}
  \frac{\bar\Sigma(r)}{\beta\Sigma_{\rm crit}}
  &=\frac{a^2M(\rvmax)}{A\pi\rvmax^2\beta\Sigma_{\rm crit}}x^2\times\notag\\
  &\int_0^xudu\int_0^\infty \frac{dv}{(u^2+v^2)^{1/2}(1+(u^2+v^2)^{1/2})^2}=1 \label{eqn:rcrit}
\end{align}
where $A=\ln(1+a)-a/(1+a)$, $a=2.16$, and $x=R_{\beta\Sigma_{\rm crit}}/\rnfw$. 

\par
\begin{figure}
    \centering
    \includegraphics[width=0.99\columnwidth]{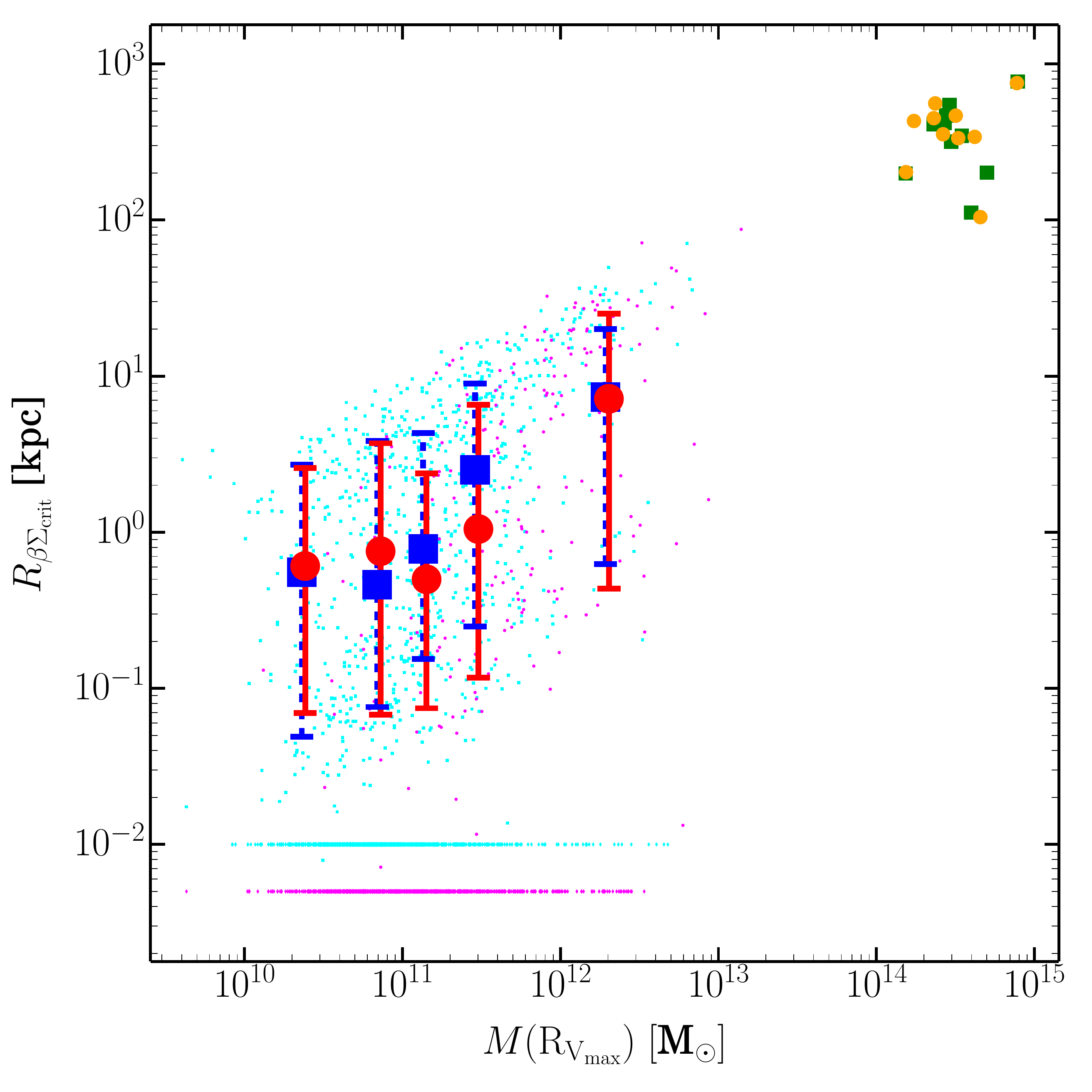}
    \caption{The radius of critical surface density where $\Sigma(R)=\beta\Sigma_{\rm crit}$ as a function of subhalo mass. Error bars, markers, colours and line styles are the same as in \Figref{fig:subhaloprops}. We also show the WDM \& CDM clusters as orange circles and green squares respectively. If a (sub)halo's $\bar\sigma$ never exceeds the criterion, the critical radius is set to 0.005 and 0.01 kpc for WDM \& CDM haloes respectively.}
    \label{fig:subhalorcrit}
\end{figure}

The resulting distribution of critical radii is shown in \Figref{fig:subhalorcrit}. In some instances, a (sub)halo is simply not dense enough. These objects have their $R_{\beta\Sigma_{\rm crit}}$ set to some low values for this plot and are ignored when calculating the median and quantiles. At first glance, it appears that the distributions don't differ significantly. Large (sub)haloes have larger critical radii. Both also show an apparent bimodal distribution, concentrated subhaloes with large critical radii and subhaloes that are being tidally disrupted resulting in the population with tiny critically radii that extends down to subhaloes that never meet the criterion outlined in \Eqref{eqn:rcrit}. The more diffuse nature of WDM subhaloes appears to shows up as a slightly smaller critical radius for well resolved subhaloes with $M(\rvmax)\sim10^{11}~\Msun$, however recall that here we are not interested in density itself but the radius at which the object is dense enough. This figure illustrates that as a result of CDM subhaloes being more dense, more of them have non-negligible critical radii. Naively, we would expect the simple fact that the greater number of subhaloes in CDM clusters would result in CDM clusters having larger lensing cross-sections. 

\par
\begin{figure}
    \centering
    \includegraphics[width=0.99\columnwidth]{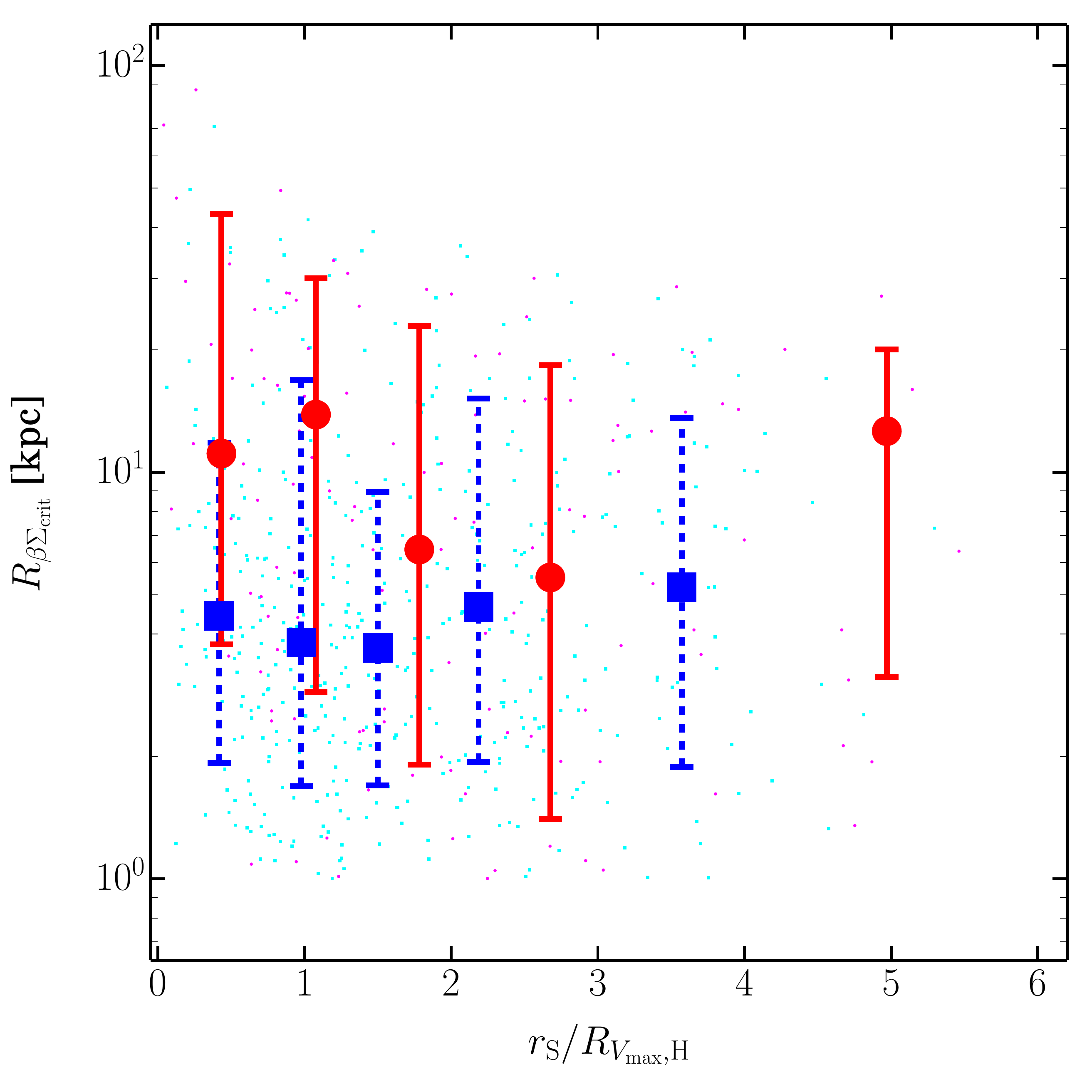}
    \caption{How substructure contributes to lensing signal. Sub-selection of subhaloes with large $R_{\beta\Sigma_{\rm crit}}$ and the radial distance to halo centre. Here the radial bins are chosen to ensure equal number of data points in each bin separately for each cosmology. Error bars, markers, colours and line styles are the same as in \Figref{fig:subhaloprops}.}
    \label{fig:subhaloradrcrit}
\end{figure}

However, if we limit our analysis to subhaloes that can significantly affect the apparent cross-section of a lensing cluster, that is those with $R_{\beta\Sigma_{\rm crit}}\gtrsim1$~kpc and examine their spatial distribution, the picture changes. Note that for both cosmologies, only $\approx10\%$ of the subhaloes residing in the cluster contribute to its strong lensing signal. WDM clusters have $\approx2.5$ times fewer subhaloes with non-negligible cross-sections as their CDM counterparts. We show the distribution of these subhaloes in \Figref{fig:subhaloradrcrit}, where we binned the data first in radial distance so that each bin contains the same number of data points, then caculate the median and quantiles of all subhaloes that lie in a given bin. This figure clearly shows that WDM subhaloes within $\lesssim2 R_{V_{\rm max},{\rm H}}$ typically have much larger critical radii compared to their CDM counterparts, particularly subhaloes at $0.5\rvmax$. The fact that the WDM data points are at larger distances than the corresponding CDM data point show WDM subhaloes are in a more radially extended distribution. 
These subhaloes are not so close to the central core of their host halo to be significantly tidally disrupted but lie at radii of $\sim200-1000$~kpc. 

The higher WDM cross-sections, which can be a factor of 2 times larger, combined with their more extended radial distribution means that these WDM subhaloes will significantly increase the Einstein radius associated with the host cluster. Despite the fact that CDM clusters contain more subhaloes with non-negligible critical radii, the cumulative effect of a few subhaloes with $R_{\beta\Sigma_{\rm crit}}\sim1$ distributed around the host is not as pronounced as that of an individual mass concentration with $R_{\beta\Sigma_{\rm crit}}\sim10$ in regards to the cross-section \cite[see][for a discussion of the effects of small substructures in numerical simulations]{xu2009}. Moreover, both cosmologies have similar number of massive subhaloes as seen in \Figref{fig:subvmaxcdf} so the excess of CDM smaller subhaloes does not play a major role \cite[see][which shows the effect of including substructure on the lensing profile in Fig. 12]{mahdi2014a}. Finally, the more extended spatial distribution of massive subhaloes with large critical radii will tend to stretch the radial arcs produced by a cluster in the WDM model relative to the CDM one.

\par
\begin{figure}
    \centering
    \includegraphics[width=0.99\columnwidth]{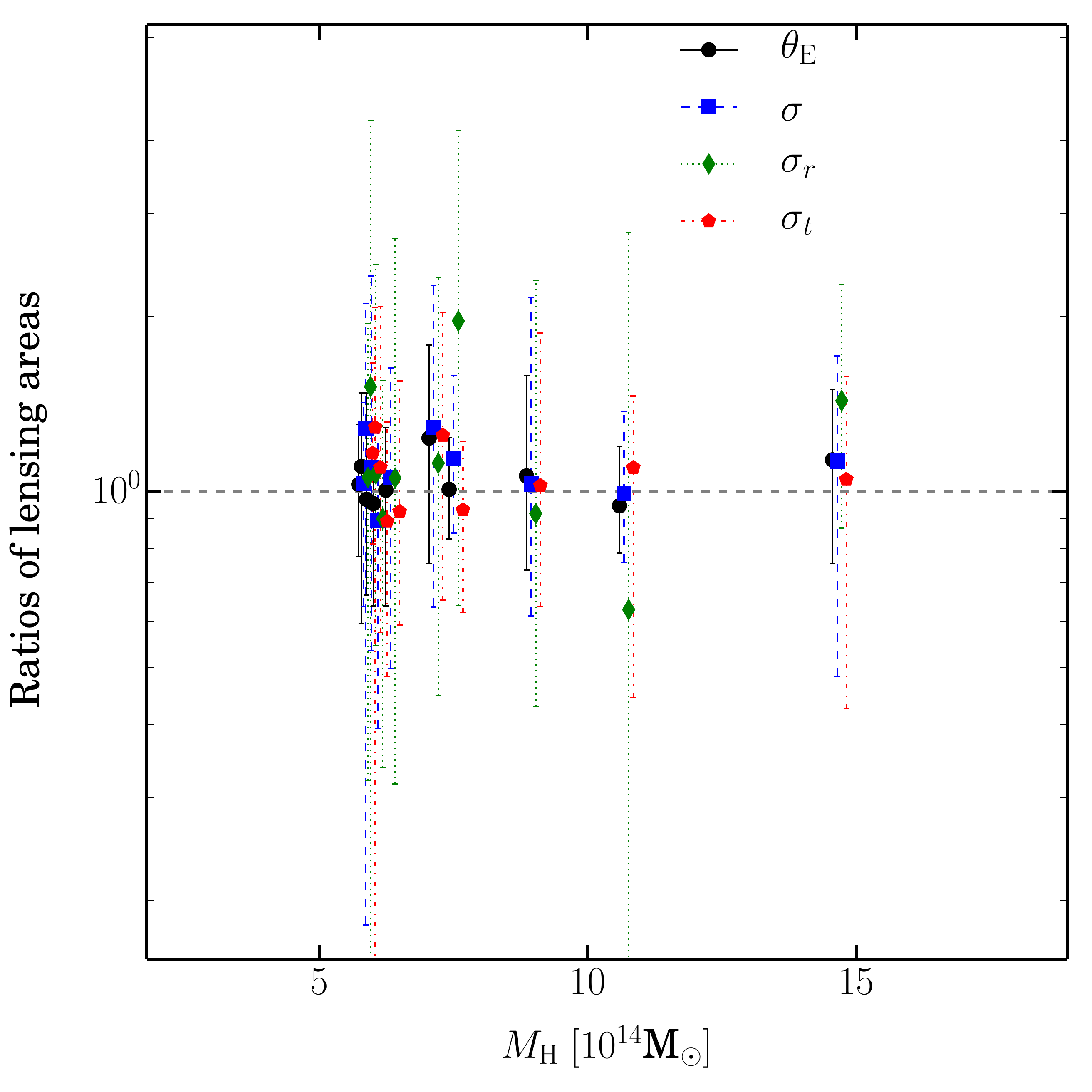}
    \caption{Ratio of WDM to CDM of several lensing quantities for each cluster as a function of cluster mass: the Einstein radius, $\theta_{\rm E}$, the total lensing cross-section, $\sigma$ and the radial and tangential cross-sections, $\sigma_{r}$ \& $\sigma_{t}$ respectively. Data points are ratio of the median values and error bars indicate the $16\%$ \& $84\%$ quantiles. We also show a dashed horizontal line at $y=1$ to guide the eye.}
    \label{fig:lensingcrossections}
\end{figure}
\begin{figure}
    \centering
    \includegraphics[width=0.99\columnwidth]{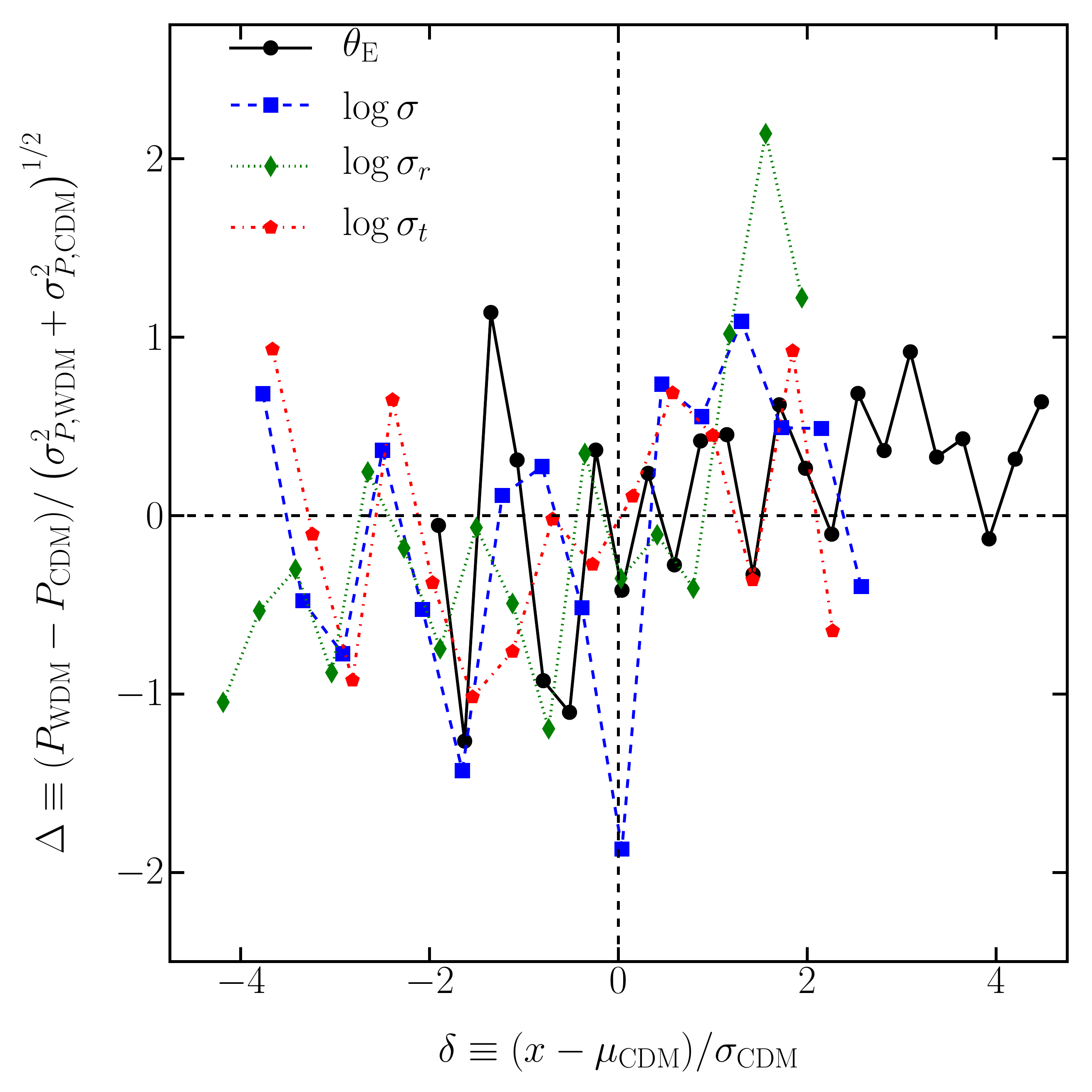}
    \caption{Difference in the probability distribution functions between WDM and CDM of several lensing quantities: the Einstein radius, $\theta_{\rm E}$, the total lensing cross-section, $\sigma$ and the radial and tangential cross-sections, $\sigma_{r}$ \& $\sigma_{t}$ respectively. Data points are the difference in the probability weighted by the error in the estimated probabilities. To plot all quantities on a single plot the ordinate is the deviation of the lensing quantity from the mean value in the CDM cosmology scaled by the standard deviation in this cosmology. We also show a dashed horizontal line at $\Delta=0$ and a vertical dashed line at $\delta=0$ to guide the eye.}
    \label{fig:lensingpdfs}
\end{figure}

The lensing signature resulting from the fact that WDM clusters have larger substructures at greater distances is summarised in figures \ref{fig:lensingcrossections} \& \ref{fig:lensingpdfs}. Here, we present two lensing quantities, the Einstein radius, $\theta_{\rm E}$ and the lensing cross-section $\sigma$. We briefly summarise the physical meaning of these quantities, for more details of how these were calculated see our companion paper \cite{mahdi2014a}. The Einstein radius is defined as the size of the tangential critical line where $\Sigma/\Sigma_{\rm crit}=1$. This is only a circular area for axisymmetric smooth mass distributions, however it is a useful quantity to calculate the angular size of a gravitational lens. The cross-section for giant arcs is defined as the area on the source plane where a source must be located in order to be lensed as a giant arc, again a measure of the area of a gravitational lens. This cross-section can be further subdivided into regions where radial arcs dominate to those where tangential arcs, like an Einstein ring, dominate. 

%\begin{comment}
\par
Figure \ref{fig:lensingcrossections} shows the median ratio and the quantiles, which are determined by taking many different lines-of-sight for each cluster. We see first that for an individual cluster, a different line-of-sight can {\em significantly} change its lensing profile. For instance the most massive cluster has has a $\theta_{\rm E}$ ratio that along some lines-of-sight is $\approx1.5$ and along others is $\approx0.75$. The amount of variation is due to the the triaxial shape of dark matter haloes and the highly anisotropic distribution of subhaloes with significant lensing cross-sections. We also see that most WDM clusters have on average marginally bigger $\theta_{\rm E}$ with the distribution skewed to larger Einstein radii. Similarly, WDM have higher total cross-sections, though the quantiles are broader and centred around unity indicating that the difference between WDM and CDM clusters is not very pronounced. However, the distribution of either strongly radial or tangential arcs is quite different. On average, WDM clusters have significantly larger radial cross-sections, whereas the cross-section for tangential arcs does not have a strong bias and in fact the quantiles are skewed to smaller values. There is only one cluster for which the trend of higher $\sigma_r$ and similar or smaller $\sigma_t$ is significantly reversed, however the variation in the ratio varies greatly depending on the line-of-sight. 
%\end{comment}

\par
The significance in the different lensing characteristics of WDM halos is seen by comparing the probability distribution functions (pdfs) shown in \cite{mahdi2014a}, which is estimated by combining the results from all clusters and lines-of-sight. In figure \ref{fig:lensingpdfs}, we show the significance of the difference between WDM and CDM, $\Delta$, by weighting the difference by the error, estimated using bootstrap re-sampling. To show the differences for all 4 lensing quantities on a single plot, we scale each cross-section by the standard deviation of the CDM distribution to represent the deviation from the mean value, $\delta$. This figure shows that $\theta_{\rm E}$, $\sigma$, and $\sigma_t$ in WDM cosmologies have an excess for large cross-section values, that is $\Delta>0$ for $\delta>0$, though given the variation, the excess is probably only significant to $\sim\tfrac{1}{2}\sigma$. More importantly, $\sigma_r$ is strongly skewed to larger values in WDM models, with excess for $\delta>0$ significant at $\gtrsim1\sigma$ level.  
%\end{comment}

\section{Discussion \& Conclusion}\label{sec:discussion} 
We have studied the evolution and properties of clusters in WDM cosmologies. We find that the removal of small-scale density perturbations to due free-streaming of dark matter particles that arises in WDM models has far reaching implications even at scales well above the dampening scale. One major consequence is that in WDM cosmologies, not only is structure formation suppressed for haloes below the dampening scale, but haloes gain mass via smooth mass accretion at much faster rates than their CDM counterparts. This enhanced smooth accretion rate is still present even at cluster scales, which are at mass scales 3 orders of magnitude larger than the dampening scale. The merger history shows clearly that WDM haloes undergo far fewer mergers than CDM haloes do, though there does not appear to be a significant trend in the reduction of major mergers. The net result of fewer mergers is clusters contain fewer substructures. These substructures are more extended and less dense than their CDM counterparts due to the higher smooth accretion rates of their progenitor haloes. Furthermore, due to their more fragile nature, they are more susceptible to tidal disruption resulting in a more radially extended distribution around their host halo. 

\par
The signatures of WDM can be observed not only in the galaxy distribution but in the gravitational lensing profile. It is often assumed that since in warm dark matter haloes contain few substructures, which in general should enhance the lensing efficiency, ``warm'' haloes should be less efficient gravitational lenses that their ``cold'' brethren.  This folk lore is wrong, WDM haloes have {\em higher} lensing efficiencies than their CDM counterparts as a result of the lensing profile being dominated by physically extended subhaloes with large critical radii. Of greater significance, WDM clusters have significantly higher cross-sections for radial arcs and marginally lower for tangential arcs relative to CDM clusters due to the more radially extended distribution of these large substructures in WDM clusters. 

\par
True, the greatest effect on the lensing profile of a cluster is its orientation to our line-of-sight due to its triaxial shape and anisotropic distribution of large subhaloes. The lensing difference observed in our companion paper, \cite{mahdi2014a}, is based on 10 clusters viewed along 150 different lines-of-sight, which at first glance amounts to a sample of 1500 clusters. However, the lensing profile will does not change drastically from one line-of-sight to another if the angle between them is small. Effectively, only lines-of-sight that are separated by angles of $\gtrsim45^{\circ}$, where $\theta_{\rm E}$ changes by $\lesssim25\%$, can be considered as arising from a different cluster. Each cluster mimic 7 different clusters and our sample effectively contains $70$ clusters. Therefore, a sample of $\approx100$ clusters should average out the variation due to differences in shape and substructure distribution and show whether there is an excess in inferred radial arc cross-section relative to the CDM prediction.

\par
The amount of enhancement naturally depends on the WDM model in question. We have only explored a single simple WDM model, that of a thermally produced dark matter particle with a mass of $0.5$~keV. This candidate results in a rather large free streaming scale. \cite{viel2013} analysis of the Lyman-$\alpha$ flux tends to strongly disfavour such light particles. Instead, they find that thermally produced dark matter particles have $m_{\rm DM}>3.3$~keV at the $2\sigma$ confidence level. Additionally, particles with $m_{\rm DM}\lesssim1$~keV would likely suppress structure formation at scales such that there would be {\em too few} satellite galaxies at Galactic scales \cite[e.g.][]{lovell2013,schneider2013b}. However, this does not invalidate our results as the gravitational lensing signature is present at scales that are {\em several orders of magnitude larger} than the scale at which the growth of structure is suppressed. 

\par
Therefore, gravitational lensing and specifically radial arcs, though difficult to identify, will allow us to measure the temperature of dark matter.

\section*{Acknowledgements}
The authors thank the anonymous referee for their constructive comments. PJE is supported by the SSimPL programme and the Sydney Institute for Astronomy (SIfA), DP130100117 and DP140100198. HSM is supported by the University of Sydney's International Scholarship programme. CP is supported by DP130100117, DP140100198, and FT130100041.

%-------------------------
\pdfbookmark[1]{References}{sec:ref}
\bibliographystyle{mn2e}
%\bibliography{ref.bib}
\bibliography{wdmlesning.bbl}

\end{document}